\documentclass[aps,singlecolumn,prd,preprintnumbers,superscriptaddress,nofootinbib]{revtex4-2}
\usepackage{hyperref}
\hypersetup{colorlinks=true,linkcolor=purple,anchorcolor=blue,citecolor=blue, filecolor=blue,urlcolor=blue,bookmarksnumbered=true,
	pdfview=FitB
}
\usepackage{ragged2e}
\usepackage[justification=centering]{caption}
\usepackage{subcaption}
\usepackage{color}
\usepackage{xcolor}
\usepackage{amsmath}
\usepackage{ulem}
\usepackage{csquotes}
\colorlet{purple1}{blue!70!red}
\colorlet{darkred}{red!50!black}
\usepackage{graphicx}
\usepackage{psfrag}
\usepackage{color}
\usepackage{comment}
\usepackage{amssymb}
\usepackage{amsmath}
\usepackage{epstopdf}
\usepackage{epsf}
\usepackage{subcaption}
\usepackage{natbib}

\newcommand{\nslash}{\kern 0.2 em n\kern -0.50em /}
\newcommand{\kslash}{\kern 0.2 em k\kern -0.45em /}
\newcommand{\lslash}{\kern 0.2 em l\kern -0.50em /}
\newcommand{\pslash}{\kern 0.2 em p\kern -0.50em /}
\newcommand{\Sslash}{\kern 0.2 em S\kern -0.50em /}
\newcommand{\Pslash}{\kern 0.2 em P\kern -0.50em /}
\newcommand{\Dslash}{\kern 0.2 em D\kern -0.65em /\kern 0.15em}

\newcommand{\be}{\begin{eqnarray}}
\newcommand{\ee}{\end{eqnarray}}

\begin{document}

\title{Gravitational transverse momentum distribution of proton}

\author{Kauship Saha}
\email{kauships24@iitk.ac.in}
\affiliation{Department of Physics, Indian Institute of Technology Kanpur, Kanpur-208016, India}

\author{Dipankar~Chakrabarti}
	\email{dipankar@iitk.ac.in} 
	\affiliation{Department of Physics, Indian Institute of Technology Kanpur, Kanpur-208016, India}  

\author{Asmita~Mukherjee}
	\email{asmita@phy.iitb.ac.in} 
	\affiliation{Department of Physics, Indian Institute of Technology Bombay, Powai, Mumbai 400076, India}

\begin{abstract}
We present the first  study of quark gravitational transverse-momentum distributions within the light-front quark--diquark model (LFQDM) inspired by the soft-wall AdS/QCD framework. We derive analytical expressions for the six unpolarized (T-even) gravitational transverse-momentum-dependent distributions (gravitational--TMDs) for up and down quarks within the model and compute the corresponding gravitational parton distribution functions (gravitational--PDFs). We further verify that these unpolarized gravitational--TMDs satisfy the model-independent relations with quark TMDs. In addition, we explore the connection of gravitational TMDs with the transverse isotropic pressure and shear-force distributions in momentum space, as well as with the average longitudinal momentum carried by up and down quarks within the model. 
\end{abstract}

\date{\today}

\maketitle
\section{Introduction}

Understanding the internal structure of the proton in terms of quarks and gluons remains a central goal of quantum chromodynamics (QCD). Over the past several decades, a wide range of high-energy scattering experiments~\cite{AbdulKhalek:2021gbh} have been carried out to probe this internal structure. In this context, deep-inelastic scattering (DIS) experiments have played a crucial role in accessing parton distribution functions (PDFs), which describe the longitudinal momentum distributions of quarks and gluons inside the proton~\cite{COLLINS1982445, Martin:1998sq}. Complementary information on the three-dimensional structure of the proton is provided by generalized parton distributions (GPDs), which can be accessed through hard exclusive processes such as deeply virtual Compton scattering (DVCS) and deeply virtual meson production (DVMP)~\cite{Ji:1996nm, Goeke:2001tz, Belitsky:2005qn, Diehl:2003ny}. GPDs encode correlations between the longitudinal momentum of partons and their transverse spatial distributions, thereby offering insight into the spatial tomography of the proton (see the recent review~\cite{Lorce:2025aqp}). In contrast, transverse-momentum-dependent parton distributions (TMDs) describe the intrinsic transverse momentum and spin correlations of partons inside the proton (for a detailed review, see Ref.~\cite{Boussarie:2023izj}). These distributions can be probed in semi-inclusive deep-inelastic scattering (SIDIS) and Drell--Yan processes~\cite{Mulders:1995dh, Barone:2001sp, Bacchetta:2006tn, Arnold:2008kf, Anselmino:2020vlp}, which provide access to the momentum-space structure of partons. Together, PDFs, GPDs, and TMDs provide complementary information and lead to a multidimensional understanding of the proton’s internal dynamics. A unified description of these partonic distributions is provided by generalized transverse-momentum-dependent parton distributions (GTMDs)~\cite{Lorce:2013pza, Meissner:2009ww, Meissner:2008ay}, which can, in principle, be accessed through exclusive double-Drell-Yan processes and exclusive $\pi^0$ production in electron-proton scattering~\cite{Bhattacharya:2017bvs, Bhattacharya:2023yvo, Bhattacharya:2023hbq}. GTMDs are often referred to as the ``mother distributions'' because, in appropriate limits and upon integration over specific variables, they reduce to PDFs, GPDs, TMDs, and form factors, thereby providing a unified framework for describing the multidimensional structure of the proton. The Fourier transform of GTMDs with respect to the transverse momentum transfer leads to Wigner distributions~\cite{Belitsky:2003nz, Lorce:2015sqe}, which provide a phase-space description of partons inside the proton. Model-based investigations of GPDs, TMDs, GTMDs, and Wigner distributions using overlaps of light-front wave functions can be found in the literature ~\cite{More:2017zqp, Mukherjee:2015aja, Mukherjee:2014nya, Chakrabarti:2025qba, Chakrabarti:2024hwx, Chakrabarti:2023djs, Gurjar:2022rcl, Chakrabarti:2016yuw, Maji:2015vsa, Mondal:2015uha}. Also, off-forward matrix elements of the QCD energy--momentum tensor (EMT) provides access to the gravitational form factors (GFFs), originally introduced in the 1960s, which encode the mechanical properties of hadrons~\cite{Pagels:1966zza, Polyakov:2018zvc, Lorce:2018egm}. The gravitational form factors basically give how matter couples to gravity. These can be accessed indirectly for example in the DVCS scattering experiments through GPDs, as the second  moment of the GPDs are related to GFFs. Due to the extreme weakness of the gravitational interaction, a  direct measurement of GFFs in present-day collider experiments is difficult; although recently, a novel approach has been suggested in ~\cite{Hatta:2023fqc}.

Meanwhile, the transverse-momentum-dependent structure of the QCD quark energy-momentum tensor (EMT) has been introduced recently by Lorcé \textit{et al.}~\cite{Lorce:2023zzg}, leading to the formulation of gravitational transverse-momentum-dependent distributions (gravitational TMDs). The primary aim of this framework is to establish a direct connection between the EMT and transverse-momentum-dependent parton distributions (TMDs), analogous to the well-known relation between the EMT and generalized parton distributions (GPDs)~\cite{Ji:1996ek}. To construct a transverse-momentum-dependent version of the quark EMT, one must consider a bilocal and gauge-invariant generalization of the local EMT operator. However, for the kinetic (Belinfante-improved) EMT~\cite{Ji:1996ek}, the presence of the covariant derivative introduces subtleties, as it does not commute with the Wilson line required for gauge invariance. As a result, the four-momentum $k$ cannot be straightforwardly interpreted as the quark four-momentum. To address this issue, one instead employs the light-front gauge-invariant canonical EMT operator, as discussed in Ref.~\cite{Lorce:2023zzg}. This operator involves the pure-gauge covariant derivative $D^{\mu}_{\textit{pure}}=\partial^\mu-igA^{\mu}_{\textit{pure}}$~\cite{Chen:2008ag, Wakamatsu:2010cb}, which reduces to the ordinary derivative $\partial^\mu$ in the light-front gauge $A^+=0$ (with appropriate boundary conditions)~\cite{Lorce:2012ce, Hatta:2011ku, Hatta:2011zs}, thereby allowing for a consistent interpretation of the quark four-momentum $k$. One may then define a fully unintegrated EMT by considering its forward matrix element in a bilocal, gauge-invariant form. The transverse-momentum-dependent EMT is obtained by further integrating this fully unintegrated correlator over the quark light-front energy. By imposing the symmetry constraints of parity, Hermiticity, and time-reversal invariance on the EMT, the resulting TMD-EMT can be parametrized in terms of 32 independent scalar functions, referred to as quark gravitational TMDs. Among these, 10 are polarization-independent and T-even, while the remaining 22 are polarization-dependent and T-odd. Upon integrating the gravitational TMDs over the transverse momentum, one obtains the corresponding gravitational parton distribution functions (PDFs); in particular, there are ten independent gravitational PDFs. These distributions encode detailed information about the partonic energy-momentum structure of the nucleon and are related to its mechanical properties, such as transverse pressure and stress distributions, as well as the average energy carried by quarks inside the nucleon, as discussed in Ref.~\cite{Lorce:2023zzg}.

At present, most studies of hadron structure based on the QCD energy-momentum tensor are formulated in terms of gravitational form factors (GFFs)~\cite{Lorc2025, Won:2025dgc, Amor-Quiroz:2023rke, Burkert:2023wzr, Lorce:2021xku, Ji:2025qax, Fujii:2025aip, Burkert:2018bqq, Polyakov:2018zvc, Shanahan:2018nnv, Hackett:2023rif, Kumericki:2019ddg, Pefkou:2021fni, Metz:2020vxd, Freese:2021czn, Polyakov:2018exb, Tong:2021ctu, Chakrabarti:2020kdc, Yao:2024ixu, Choudhary:2022den, Nair:2024fit, More:2021stk, Dehghan:2025ncw, Sain:2025kup, Chakrabarti:2021mfd, Chakrabarti:2015lba, Kumar:2017dbf, More:2023pcy, Choudhary:2022den, Sain:2025kup, Chakrabarti:2021mfd, Chakrabarti:2015lba, Cao:2026jzm}. In contrast, gravitational transverse-momentum-dependent distributions have not yet been systematically investigated, either within phenomenological models or in lattice QCD calculations. In this work, we present a first study of the unpolarized quark gravitational TMDs within the light-front quark--diquark model~\cite{Maji:2016yqo} and compute the corresponding gravitational PDFs, thereby taking an initial step toward understanding these distributions in a model framework. The light-front wave functions employed in this model are based on the soft-wall AdS/QCD prediction~\cite{Maji:2016yqo, Lepage:1980fj, PhysRevD.77.056007}. This model has previously been applied successfully to the study of several proton properties, including PDFs, TMDs, GPDs, GTMDs, Wigner distributions, gravitational form factors, and spin asymmetries~\cite{Chakrabarti:2017teq, Maji:2017bcz, Maji:2017ill, Kumar:2017dbf, Chakrabarti:2019wjx, Maji:2022tog, Maji:2017wwd, Maji:2017zbx, Maji:2016yqo, Mondal:2016xsm},  as well as investigations of model dependent relations between GPDs and TMDs~\cite{Gurjar:2021dyv}. Using the gravitational TMDs obtained in this work, we further compute the transverse pressure, shear distributions, and the average longitudinal momentum  carried by up and down quarks within this model.

The paper is organized as follows: In Sec.~\ref{sec:LIGHT-FRONT QUARK-DIQUARK MODEL}, we briefly review the light-front quark-diquark model. In Sec.~\ref{sec:GRAVITATIONAL TRANSVERSE MOMENTUM DEPENDENT PARTON DISTRIBUTION}, we define the gravitational transverse-momentum-dependent correlator, present its parametrization in terms of gravitational TMDs, and derive the analytical expressions for the gravitational TMDs and the corresponding gravitational PDFs within the model. In Sec.~\ref{sec:Numerical result and Discussion}, we present and discuss the numerical results. Finally, in Sec.~\ref{sec:conclusion}, we summarize our main findings and conclude.

\section{LIGHT-FRONT QUARK-DIQUARK MODEL}
\label{sec:LIGHT-FRONT QUARK-DIQUARK MODEL}

In this section, we briefly summarize  the light front quark-diquark model(LFQDM) proposed in Ref.~\cite{Maji:2016yqo}. In this model, the proton state is expressed as a superposition of an isoscalar-scalar diquark singlet state \( |u S^0\rangle \), an isoscalar-vector diquark state \( |u A^0\rangle \), and an isovector-vector diquark state \( |d A^1\rangle \)~\cite{Bacchetta:2008af,Rodrigues:1995eu}, consistent with spin-flavor \textit{SU}(4) symmetry. The proton state can be written as
\begin{equation}
|P;\pm\rangle 
= C_S\, |u S^0\rangle^\pm 
+ C_V\, |u A^0\rangle^\pm 
+ C_{VV}\, |d A^1\rangle^\pm .
\label{eg:proton-state}
\end{equation}
where $S$ and $A$ denote the scalar and vector diquark configurations, respectively, and the superscript specifies their isospin. The coefficients $C_i$ corresponding to the scalar and vector diquark components were determined in Ref.~\cite{Maji:2016yqo}. Their numerical values are given by $C_S^2 = 1.3872$, $C_V^2 = 0.6128$, and $C_{VV}^2 = 1$.
\vspace{0.5em}

We use the light-front convention $x^{\pm}=x^{0}\pm x^{3}$. We choose a frame in which the transverse momentum of the proton vanishes, so that the proton momentum is given by
$P=\left(P^{+},\, \frac{M^2}{P^+},\, \mathbf{0}_{\perp}\right)$.
In this frame, the momentum of the struck quark is
$k \equiv \left(xP^{+},\, \frac{m^2+|\mathbf{k_{\perp}}|^2}{xP^+},\, \mathbf{k}_{\perp}\right)$,
while the momentum of the spectator diquark is
$P_X \equiv \left((1-x)P^{+},\, P_X^{-},\, -\mathbf{k}_{\perp}\right)$.
Here, $x=k^{+}/P^{+}$ is the longitudinal momentum fraction carried by the struck quark.
The two-particle Fock-state expansion for $J^{z}=\pm \tfrac{1}{2}$ can be written, for the scalar diquark case, as

\begin{equation}
\lvert u S \rangle^\pm
= \int \frac{dx\,d^2\mathbf{k}_{\perp}}{2(2\pi)^3\sqrt{x(1-x)}}\,
\Big[
\psi^{\pm(u)}_{+}(x,\mathbf{k}_{\perp})\,\lvert +\tfrac{1}{2}s;xP^+,\mathbf{k}_{\perp}\rangle
+\psi^{\pm(u)}_{-}(x,\mathbf{k}_{\perp})\,\lvert -\tfrac{1}{2}s;xP^+,\mathbf{k}_{\perp}\rangle
\Big].
\label{eq:scalar part fock state}
\end{equation}

and the light front wave-functions(LFWFs) with spin-0 diquark, for $J=\pm1/2$, are given by~\cite{Lepage:1980fj}  

\begin{equation}
\begin{aligned}
\psi^{+(u)}_{+}(x,\mathbf{k}_{\perp}) &= N_S\,\phi^{(u)}_1(x,\mathbf{k}_{\perp}),\\[4pt]
\psi^{+(u)}_{-}(x,\mathbf{k}_{\perp}) &= -\,N_S\,\frac{k^1 + i k^2}{x M}\,\phi^{(u)}_2(x,\mathbf{k}_{\perp}),\\[4pt]
\psi^{-(u)}_{+}(x,\mathbf{k}_{\perp}) &= N_S\,\frac{k^1 - i k^2}{x M}\,\phi^{(u)}_2(x,\mathbf{k}_{\perp}),\\[4pt]
\psi^{-(u)}_{-}(x,\mathbf{k}_{\perp}) &= N_S\,\phi^{(u)}_1(x,\mathbf{k}_{\perp}).
\label{eq:Scalar part wavefunction}
\end{aligned}
\end{equation}

where $\lvert \lambda_q \lambda_s; xP^+, \mathbf{k}_\perp \rangle$ denotes a two-particle state consisting of a struck quark with helicity $\lambda_q$ and a scalar diquark with helicity $\lambda_s = s$. The spin--$0$ singlet diquark helicity is denoted by $s$ to distinguish it from the triplet diquark. The state with a spin-$1$ diquark is given by~\cite{Ellis:2008in}
\begin{equation}
\begin{split}
\lvert \beta A \rangle^\pm
= \int \frac{dx\,d^2\mathbf{k}_{\perp}}{2(2\pi)^3}\,\frac{1}{\sqrt{x(1-x)}}\,
\Big[ 
&\psi^{\pm(\beta)}_{++}(x,\mathbf{k}_{\perp})\,\lvert +\tfrac{1}{2}+1;xP^+,\mathbf{k}_{\perp}\rangle
+ \psi^{\pm(\beta)}_{-+}(x,\mathbf{k}_{\perp})\,\lvert -\tfrac{1}{2}+1;xP^+,\mathbf{k}_{\perp}\rangle \\
&+ \psi^{\pm(\beta)}_{+0}(x,\mathbf{k}_{\perp})\,\lvert +\tfrac{1}{2}0;xP^+,\mathbf{k}_{\perp}\rangle
+ \psi^{\pm(\beta)}_{-0}(x,\mathbf{k}_{\perp})\,\lvert -\tfrac{1}{2}0;xP^+,\mathbf{k}_{\perp}\rangle \\
&+ \psi^{\pm(\beta)}_{+-}(x,\mathbf{k}_{\perp})\,\lvert +\tfrac{1}{2}-1;xP^+,\mathbf{k}_{\perp}\rangle
+ \psi^{\pm(\beta)}_{--}(x,\mathbf{k}_{\perp})\,\lvert -\tfrac{1}{2}-1;xP^+,\mathbf{k}_{\perp}\rangle
\Big].
\label{eq:Vector_part fock state}
\end{split}
\end{equation}
Here, $\lvert \lambda_q \lambda_D; xP^+, \mathbf{k}_\perp \rangle$ represents a two-particle state composed of a struck quark with helicity $\lambda_q = \pm \tfrac{1}{2}$ and a vector diquark with helicity $\lambda_D = \pm 1, 0$, corresponding to the triplet diquark states. The associated light-front wave functions (LFWFs) for $J = +\tfrac{1}{2}$ are given by
\begin{equation}
\begin{aligned}
\psi^{+(\beta)}_{++}(x,\mathbf{k}_\perp)
&=
N^{(\beta)}_1 \sqrt{\frac{2}{3}}
\left( \frac{k^1 - i k^2}{xM} \right)
\phi^{(\beta)}_2(x,\mathbf{k}_\perp), \\
\psi^{+(\beta)}_{-+}(x,\mathbf{k}_\perp)
&=
N^{(\beta)}_1 \sqrt{\frac{2}{3}}
\phi^{(\beta)}_1(x,\mathbf{k}_\perp), \\
\psi^{+(\beta)}_{+0}(x,\mathbf{k}_\perp)
&=
-\,N^{(\beta)}_0 \sqrt{\frac{1}{3}}
\phi^{(\beta)}_1(x,\mathbf{k}_\perp), \\
\psi^{+(\beta)}_{-0}(x,\mathbf{k}_\perp)
&=
N^{(\beta)}_0 \sqrt{\frac{1}{3}}
\left( \frac{k^1 + i k^2}{xM} \right)
\phi^{(\beta)}_2(x,\mathbf{k}_\perp), \\
\psi^{+(\beta)}_{+-}(x,\mathbf{k}_\perp)
&= 0, \\
\psi^{+(\beta)}_{--}(x,\mathbf{k}_\perp)
&= 0 .
\label{eq: vector wavefunction for up}
\end{aligned}
\end{equation}
Similarly, for $J = -\tfrac{1}{2}$, the LFWFs are
\begin{equation}
\begin{aligned}
\psi^{-(\beta)}_{++}(x,\mathbf{k}_\perp)
&= 0, \\
\psi^{-(\beta)}_{-+}(x,\mathbf{k}_\perp)
&= 0, \\
\psi^{-(\beta)}_{+0}(x,\mathbf{k}_\perp)
&=
N^{(\beta)}_0 \sqrt{\frac{1}{3}}
\left( \frac{k^1 - i k^2}{xM} \right)
\phi^{(\beta)}_2(x,\mathbf{k}_\perp), \\
\psi^{-(\beta)}_{-0}(x,\mathbf{k}_\perp)
&=
N^{(\beta)}_0 \sqrt{\frac{1}{3}}
\phi^{(\beta)}_1(x,\mathbf{k}_\perp), \\
\psi^{-(\beta)}_{+-}(x,\mathbf{k}_\perp)
&=
-\,N^{(\beta)}_1 \sqrt{\frac{2}{3}}
\phi^{(\beta)}_1(x,\mathbf{k}_\perp), \\
\psi^{-(\beta)}_{--}(x,\mathbf{k}_\perp)
&=
N^{(\beta)}_1 \sqrt{\frac{2}{3}}
\left( \frac{k^1 + i k^2}{xM} \right)
\phi^{(\beta)}_2(x,\mathbf{k}_\perp).
\label{eq: vector wavefunction for down}
\end{aligned}
\end{equation}

having flavor index $\beta = u,d$, where $N_s$, $N_{0}^{(\beta)}$, and $N_{1}^{(\beta)}$ are normalization constants whose values are listed in Table~\ref{tab:Normalizationconstant}. The light-front wave functions $\phi_{i}^{(\beta)}(x, \mathbf{k}_{\perp})$ are taken to be a modified form of the soft-wall AdS/QCD prediction and are given by
\begin{equation}
\phi^{(\beta)}_i(x,\mathbf{k}_\perp)
=
\frac{4\pi}{\kappa} \,\sqrt{\frac{\log(1/x)}{(1-x)}}
\,x^{a_i^{\beta}} (1-x)^{b_i^{\beta}}
\exp\!\left[
-\delta^{\beta}\,\frac{\mathbf{k}_\perp^{\,2}}
{2\kappa^{2}}
\frac{\log(1/x)}{(1-x)^{2}}
\right],
\qquad (i=1,2).
\label{eq: Ads/ACD wavefunction}
\end{equation}

\begin{table}[h]
\centering
\renewcommand{\arraystretch}{1.5} 
\setlength{\tabcolsep}{12pt}       
\caption{Values of normalization constants $N_i$ corresponding to both u and d quarks.}
\label{tab:Normalizationconstant}
\begin{tabular}{c c c c}
\hline
$\nu$ & $N_{S}^{(\beta)}$ & $N_0^{(\beta)}$ & $N_1^{(\beta)}$ \\
\hline
$u$ & 2.0191 & 3.2050 & 0.9895 \\
$d$ & 0      & 5.9423 & 1.1616 \\
\hline
\end{tabular}
\end{table}

The wave functions $\phi_{i}^{(\beta)}$ $(i=1,2)$ reduce to the AdS/QCD prediction~\cite{deTeramond:2011aml,PhysRevD.77.056007} when the parameters satisfy $a_{i}^{\beta}=b_{i}^{\beta}=0$ and $\delta^{\beta}=1.0$. Throughout this work, we use the AdS/QCD scale parameter $\kappa = 0.4~\mathrm{GeV}$, as determined in Refs.~\cite{Chakrabarti:2013dda,Chakrabarti:2013gra}. The constituent quark mass $m$ and the proton mass $M$ are taken to be $0.055~\mathrm{GeV}$ and $0.938~\mathrm{GeV}$, respectively, following Ref.~\cite{Chakrabarti:2019wjx}. The parameters $a_{i}^{\beta}$ and $b_{i}^{\beta}$ are fitted at the model initial scale $\mu_0 = 0.313~\mathrm{GeV}$ using the Dirac and Pauli form factors. At this scale, the parameter $\delta^{\beta}$ is taken to be unity for both up and down quarks~\cite{Maji:2016yqo}. The model parameters for both up and down quarks used in this work, evaluated at the initial scale $\mu_0 = 0.313~\mathrm{GeV}$, are listed in Table~\ref{tab:fitted_parameters}.

\begin{table}[ht]
\centering
\caption{The fitted parameters for $u$ and $d$ quarks.}
\label{tab:fitted_parameters}
\begin{ruledtabular}
\begin{tabular}{c c c c c c}
$\beta$ 
& $a_1^{\beta}$ 
& $b_1^{\beta}$ 
& $a_2^{\beta}$ 
& $b_2^{\beta}$ 
& $\delta^{\beta}$ \\
\hline
$u$ 
& $0.280 \pm 0.001$ 
& $0.1716 \pm 0.0051$ 
& $0.84 \pm 0.02$ 
& $0.2284 \pm 0.0035$ 
& $1.0$ \\

$d$ 
& $0.5850 \pm 0.0003$ 
& $0.7000 \pm 0.0002$ 
& $0.9434^{+0.0017}_{-0.0013}$ 
& $0.64^{+0.0082}_{-0.0022}$ 
& $1.0$ \\
\end{tabular}
\end{ruledtabular}
\end{table}

\section{GRAVITATIONAL TRANSVERSE MOMENTUM DEPENDENT PARTON DISTRIBUTION}
\label{sec:GRAVITATIONAL TRANSVERSE MOMENTUM DEPENDENT PARTON DISTRIBUTION}

Quark gravitational TMDs encode a wealth of information about the mechanical properties of hadrons, such as transverse pressure and shear forces. In this work, we focus exclusively on unpolarized gravitational TMDs and extract the average longitudinal momentum of the quark as well as the transverse pressure for an unpolarized proton in the model.
 
The fully unintegrated light-front energy-momentum tensor for a quark is defined as~\cite{Lorce:2023zzg}
\begin{equation}
\Theta_q^{\mu\nu}(P,k,N,S;\eta)
= \frac{1}{2}
\int \frac{d^4z}{(2\pi)^4}
\, e^{ik\cdot z}\,
i\partial_z^\nu
\left.
\langle P,S|
\bar{\psi}(-\tfrac{z}{2})\,\gamma^\mu\,
\mathcal{W}(-\tfrac{z}{2},\tfrac{z}{2}|n)\,
\psi(\tfrac{z}{2})
|P,S\rangle
\right.
\label{eq:unintegeratedTMD}
\end{equation}
where, $\lvert P,S \rangle$ denotes the hadron state with four-momentum $P$,  covariant-spin $S$, and $k$ is the four-momentum of the quark. The TMD energy-momentum tensor is obtained by integrating over the quark light-front energy, i.e.,
\begin{equation}
\begin{aligned}
\mathcal{T}_q^{\mu\nu}(P,x,\boldsymbol{k}_\perp,N,S;\eta)
&= \int dk^-\, \Theta_q^{\mu\nu}(P,k,N,S;\eta)
\\[0.2em]
&= \frac{1}{2}
\int \frac{dz^-\, d^2\boldsymbol{z}_\perp}{2(2\pi)^3}
\, e^{ik\cdot z}\,
i\partial_z^\nu
\left.
\langle P,S|
\psi(-\tfrac{z}{2})\,\gamma^\mu\,
\mathcal{W}(-\tfrac{z}{2},\tfrac{z}{2}|n)\,
\psi(\tfrac{z}{2})
|P,S\rangle
\right|_{z^+=0}.
\end{aligned}
\label{eg: TMD-EMT}
\end{equation}

This TMD energy-momentum tensor can be interpreted as a three-dimensional distribution of the quark energy-momentum tensor in momentum space. Gauge invariance is ensured by the inclusion of the Wilson line $\mathcal{W}$ between the quark field operators, extending from the space-time point $-\tfrac{z}{2}$ to $+\tfrac{z}{2}$ along the lightlike direction $n$. This Wilson line is invariant under the rescaling $n \to \alpha n$ with $\alpha > 0$. Consequently, the correlator depends only on the rescaling-invariant four-vector
\begin{equation}
N = \frac{M^2\, n}{P \cdot n}\, .
\end{equation}
The light-cone vector is defined as $n^\mu = (0,\, \eta,\, \mathbf{0}_\perp)$, where the parameter $\eta = \mathrm{sign}(n^0)$ specifies the direction of the Wilson line. In particular, $\eta = +1$ corresponds to a future-pointing Wilson line, while $\eta = -1$ corresponds to a past-pointing one.
\subsection{Parameterization in terms of gravitational TMDs}

The  parametrization of  the transverse momentum-dependent energy-momentum tensor is based on the symmetry constraints -parity, hermiticity, and time-reversal imposed on the fully unintegrated TMD defined in Eq.~\eqref{eq:unintegeratedTMD}, as discussed in Ref.~\cite{Lorce:2023zzg}. To begin with, we express the covariant spin vector of the nucleon as
\begin{equation}
S^\mu = \frac{\lambda}{M}(P^\mu - N^\mu) + S_T^\mu .
\end{equation}
where the longitudinal polarization is denoted by $\lambda$, and the transverse polarization is given by $S_{T}^\mu=\big(0,0,\mathbf{S}_{\perp}\big)$. Following Ref.\cite{Lorce:2023zzg},  we define the transverse metric tensor and the transverse Levi-Civita tensor as:
\begin{equation}
\begin{aligned}
g_{\perp}^{\mu\nu}
&= g^{\mu\nu}
- \frac{P^\mu N^\nu + P^\nu N^\mu}{M^{2}}
+ \frac{N^\mu N^\nu}{M^{2}}, \\
\epsilon_{\perp}^{\mu\nu}
&= \frac{\epsilon^{\mu\nu\alpha\beta}\, N_\alpha P_\beta}{M^{2}} .
\end{aligned}
\end{equation}
The most general parametrization of the TMD energy-momentum tensor in Eq.~\eqref{eg: TMD-EMT} for spin-$0$ and spin-$\tfrac{1}{2}$ hadrons in terms of the available tensor structures $P^\mu$, $N^\mu$, $k_{\perp}^\mu$, $\epsilon_{\perp}^{\mu\nu}$, and $g_{\perp}^{\mu\nu}$, is given by
\begin{equation}
\small
\begin{aligned}
\mathcal{T}_{q}^{\mu\nu}
= \frac{1}{P^+}\Bigg\{&
P^\mu P^\nu a_1
+ N^\mu N^\nu a_2
+ k_{\perp}^\mu k_{\perp}^\nu a_3
+ P^\mu N^\nu a_4
+ N^\mu P^\nu a_5
+ P^\mu k_{\perp}^\nu a_6
+ k_{\perp}^\mu P^\nu a_7
\\[0.2em]
&+ N^\mu k_{\perp}^\nu a_8
+ k_{\perp}^\mu N^\nu a_9
+ M^2 g_{\perp}^{\mu\nu} a_0
\\[0.4em]
&-\frac{\epsilon_T^{k_{\perp} S_T}}{M}\Big(
P^\mu P^\nu a^{\perp}_{1T}
+ N^\mu N^\nu a^{\perp}_{2T}
+ k_{\perp}^\mu k_{\perp}^\nu a^{\perp}_{3T}
+ P^\mu N^\nu a^{\perp}_{4T}
+ N^\mu P^\nu a^{\perp}_{5T} \\
&\qquad\qquad\quad
+ P^\mu k_{\perp}^\nu a^{\perp}_{6T}
+ k_{\perp}^\mu P^\nu a^{\perp}_{7T}
+ N^\mu k_{\perp}^\nu a^{\perp}_{8T}
+ k_{\perp}^\mu N^\nu a^{\perp}_{9T}
+ M^2 g_{\perp}^{\mu\nu} a^{\perp}_{0T}
\Big)
\\[0.4em]
&-M\Big(
P^\mu \epsilon_{\perp}^{\nu S_T} a_{1T}
+ P^\nu \epsilon_{\perp}^{\mu S_T} a_{2T}
+ N^\mu \epsilon_{\perp}^{\nu S_T} a_{3T}
+ N^\nu \epsilon_{\perp}^{\mu S_T} a_{4T} \\
&\qquad\qquad\quad
+ k_{\perp}^\mu \epsilon_{\perp}^{\nu S_T} a_{5T}
+ k_{\perp}^\nu \epsilon_{\perp}^{\mu S_T} a_{6T}
\Big)
\\[0.4em]
&-\lambda \Big(
P^\mu \epsilon_{\perp}^{\nu k_{\perp}} a_{1L}
+ P^\nu \epsilon_{\perp}^{\mu k_{\perp}} a_{2L}
+ N^\mu \epsilon_{\perp}^{\nu k_{\perp}} a_{3L}
+ N^\nu \epsilon_{\perp}^{\mu k_{\perp}} a_{4L} \\
&\qquad\qquad\quad
+ k_{\perp}^\mu \epsilon_{\perp}^{\nu k_{\perp}} a_{5L}
+ k_{\perp}^\nu \epsilon_{\perp}^{\mu k_{\perp}} a_{6L}
\Big)
\Bigg\}.
\label{eg: parameterizedTMD}
\end{aligned}
\end{equation}
Here, the scalar functions $a_i(x,\mathbf{k}_{\perp}^2)$ are referred to as gravitational TMDs. For a spin-$0$ hadron, there are ten polarization-independent gravitational TMDs, denoted by $a_{0}$--$a_{9}$. For a spin-$\tfrac{1}{2}$ hadron, an additional twenty-two polarization-dependent gravitational TMDs appear. Interestingly, the same number of gravitational generalized parton distributions (GPDs) is obtained in Ref.~\cite{Lorce:2015lna}. The polarization-dependent TMDs are T-odd which changing sign under $\eta \rightarrow -\eta$, whereas the polarization-independent TMDs are T-even which remain unchanged under $\eta \rightarrow -\eta$. The functions $a_i(x,\mathbf{k}_{\perp}^2)$ depend on the longitudinal momentum fraction $x = k^+/P^+$ and the transverse momentum squared $\mathbf{k}_{\perp}^2$ of the quark. Upon integrating Eq.~\eqref{eg: parameterizedTMD} over $\mathbf{k}_{\perp}$, one obtains the corresponding gravitational PDFs. Note that all terms linear in $\mathbf{k}_{\perp}$ vanish upon integration $\int d^2\mathbf{k}_\perp\, \mathcal{T}^{\mu\nu}$, implying that there are in total ten independent gravitational PDFs.

\subsection{T-even gravitational TMDs}

In this subsection, we compute the unpolarized gravitational TMDs, which correspond to the T-even sector. The gauge-link Wilson line $\mathcal{W}_{[0,z]}$ runs along the path
$[0,0,\mathbf{0}_\perp] \rightarrow [0,1,\mathbf{0}_\perp] \rightarrow [0,1,\boldsymbol{z}_{\perp}] \rightarrow [0,z^-,\boldsymbol{z}_{\perp}]$, as discussed in Refs.~\cite{Bacchetta:2008af,Boer:2003cm}. It is known that for T-even TMDs, in the light-cone gauge $(A^{+}=0)$, the transverse  part of the Wilson line at light-cone infinity does not contribute. In this work  for the calculation of T-even gravitational TMDs also, we have taken the gauge link to be unity in the light cone gauge, and ignored this part.

%
\begin{equation}
\begin{aligned}
\mathcal{T}_q^{\mu \nu}(P,x,\boldsymbol{k}_\perp,N,S;\eta)
= \frac{1}{2}
\int \frac{dz^-\, d^2\boldsymbol{z}_\perp}{2(2\pi)^3}
\, e^{ik\cdot z}\,
i\partial_z^\nu
\left.
\langle P,S|
\bar{\psi}(0)\,\gamma^\mu\,
\psi(z)
|P,S\rangle
\right|_{z^+=0}.
\end{aligned}
\label{eq: TMD-EMT_plus}
\end{equation}

The term $\bar{\psi}(0)\gamma^{\mu}\psi(z)$ in Eq.~\eqref{eq: TMD-EMT_plus}, with $\mu = -,\, i$, corresponds to twist-3 and twist-4 operator contributions in the correlator. Within the quark–diquark model, we neglect gluonic contributions and insert the free-field light-front Fourier expansion of the quark field $\psi$ on the surface $z^+=0$. This is equivalent to what is done in the light-front constituent quark model(LFCQM) for the calculation of unpolarized T-even TMDs~\cite{Lorce:2014hxa}. The free-field light-front Fourier expansion of the quark field is given by

\begin{equation}
\psi(z^-, \mathbf{z}_{\perp})
=
\int \frac{dp^+\, d^2\boldsymbol{p}_\perp}{2p^+(2\pi)^3}
\Theta(p^+)
\sum_{r}
\left[
b^q_r(p)\, u_r(p)\,
e^{-i\frac{p^+ z^-}{2} + i \boldsymbol{p}_\perp \cdot \boldsymbol{z}_\perp}
+
d^{q\dagger}_r(p)\, v_r(p)\,
e^{i\frac{p^+ z^-}{2} - i \boldsymbol{p}_{\perp} \cdot \boldsymbol{z}_{\perp}}
\right].
\label{eq:free-quark-field}
\end{equation}

where $b^q$ and $d^{q \dagger}$ are the annihilation operator of the quark and the creation operator of the antiquark, respectively. The symbol $r$ denotes the light-front helicity of the quark, and $k$ represents the light-front four-momentum. Using Eq.~\eqref{eq:free-quark-field}, the quark contribution to the operator $\bar{\psi}(0)\gamma^{\mu}\psi(z)$ in Eq.~\eqref{eq: TMD-EMT_plus} is given by

\begin{equation}
    \bar{\psi}(0)\,\gamma^\mu\,
\psi(z)=
\int \frac{dp^+\, d^2\boldsymbol{p}_\perp}{2p^+(2\pi)^3}\,
\Theta(p^+)
\int \frac{dk^{\prime +}\, d^2\boldsymbol{k}'_\perp}{2k^{\prime +}(2\pi)^3}\, \Theta(k^{\prime +}) \sum_{r,r'}
\bar{u}_{r'}(k')
\,\gamma^\mu\,
u_{r}(p)\,
b^{q\dagger}_{r'}(k')
\,b^q_{r}(p) e^{-ip\cdot z}.
\label{eq:1st_part}
\end{equation}

From Eq.~\eqref{eq: TMD-EMT_plus}, the operator part(excluding the nucleon states) is written using Eq.~\eqref{eq:1st_part} as

\begin{equation}
\begin{aligned}
&\frac{1}{2}\int \frac{dz^-\, d^2\boldsymbol{z}_\perp}{2(2\pi)^3}\,
e^{ik\cdot z}\,
i\partial_{z}^\nu\bigg(\psi(0)\,\gamma^\mu\,\psi(z)\bigg)\bigg|_{z^+=0}
\\[2mm]
&= \frac{1}{2}
\int \frac{dp^+\, d^2\boldsymbol{p}_\perp}{2p^+(2\pi)^3}\,
\Theta(p^+) \, p^\nu
\int \frac{dk^{\prime +}\, d^2\boldsymbol{k}'_{\perp}}{2k^{\prime +}(2\pi)^3}\,
\Theta(k^{\prime +})
\\
&\quad \times
\delta(p^+ - k^+)\,
\delta^{(2)}(\boldsymbol{p}_\perp - \boldsymbol{k}_\perp)
\sum_{r,r'}
\bar{u}_{r'}(k')
\,\gamma^\mu\,
u_{r}(p)\,
b^{q\dagger}_{r'}(k')
\,b^q_{r}(p) .
\qquad \nu \neq -
\label{eq:2nd_part}
\end{aligned}
\end{equation}

By substituting Eq.~\eqref{eq:2nd_part} in the T-even gravitational TMD correlator in Eq.~\eqref{eq: TMD-EMT_plus}, we obtain 

 \begin{equation}
\mathcal{T}^{\mu\nu}_{q}(x,\mathbf{k}_{\perp}^2,S )
=
\sum_{r,r'}
\frac{\bar{u}_{r'}(k)\,\gamma^\mu\,u_{r}(k)}{2k^+}\,
\, k^\nu 
\mathcal{P}^q_{r\,r'}(x,\mathbf{k}_{\perp}^2, S) .
\label{eq:3rd_part}
\end{equation}

where the quantity $\mathcal{P}^q_{r\,r'}$ denotes the quark density matrix\cite{Lorce:2014hxa} in the space of quark light-front helicities, which is given by
\begin{equation}
\mathcal{P}^q_{r\,r'}(x,\mathbf{k}_{\perp}^2, S)
=
\frac{1}{2(2\pi)^3}
\int \frac{dk^{\prime +}\, d^2\boldsymbol{k}'_T}
     {2k^{\prime +}(2\pi)^3}
\Theta(k^{\prime +})
\,
\langle P, S|
b^{q\dagger}_{r'}(k')
\,b^q_{r}(k)
|P,S\rangle .
\label{eq:desityoperator}
\end{equation}
The trace of this matrix,
\begin{equation}
\mathcal{P}^q(x,\mathbf{k}_{\perp}^2, S)
=
\sum_{r}
\mathcal{P}^q_{r\,r}(x,\mathbf{k}_{\perp}^2,S)\,.
\label{eq:density operator}
\end{equation}
defines the quark density operator\cite{Lorce:2014hxa} evaluated in the target. Using the standard light-front spinor normalization, one finds
\begin{equation}
\bar{u}_{r'}(k)\,\gamma^\mu\,u_{r}(k)
= 2 k^\mu \,\delta_{r,r'} \, .
\label{eq:identity}
\end{equation}

By inserting Eq.~\eqref{eq:identity} in the correlator Eq.~\eqref{eq:3rd_part}, we get

\begin{equation}
\mathcal{T}_{q}^{\mu\nu}(x,\mathbf{k}_{\perp}^2,S)
=
\frac{k^\mu k^\nu}{k^+}\mathcal{P}^{q}(x,\mathbf{k}_{\perp}^2,S),
\qquad \nu \neq -
\label{eg:main_equation}
\end{equation}

This provides the most simplified expression for extracting the unpolarized gravitational TMDs. As can be seen from Eq.~\eqref{eg: parameterizedTMD}, the components with $\mu=\nu=-$ correspond to the function $a_{2}$ and $a_{4}$, while the components with $\mu=i$ and $\nu=-$ correspond to $a_{9}$. However, in Eq.~\eqref{eg:main_equation}, the index $\nu \neq -$, and therefore we focus only on extracting the remaining unpolarized gravitational-TMDs. It is important to note that these unpolarized gravitational TMDs, together with $a_{0}$, are not connected to any hadronic observable. In particular, $a_{0}=0$ when relating gravitational TMDs to standard TMDs 
\cite{Lorce:2023zzg}. Using Eq.~\eqref{eg:proton-state} in the correlator Eq.~\eqref{eq: TMD-EMT_plus}, we obtain the flavor decomposition of the correlator in terms of up and down-quark contributions as
\begin{equation}
\mathcal{T}_{u}^{\mu \nu}
= C_{S}^{2}\,\mathcal{T}_{S}^{\mu \nu}
+ C_{V}^{2}\,\mathcal{T}_{V}^{\mu \nu},
\label{eg:up_quark}
\end{equation}
\begin{equation}
\mathcal{T}_{d}^{\mu \nu}
= C_{VV}^{2}\,\mathcal{T}_{VV}^{\mu \nu},
\label{eg:d_quark}
\end{equation}

Using Eq.~\eqref{eq:scalar part fock state} in Eq.~\eqref{eq:desityoperator}, we obtain the density operator for the scalar-diquark sector in terms of overlapped representation of LFWFs as
\begin{equation}
\mathcal{P}^{S(u)}(x,\mathbf{k}_{\perp}^2; \pm)
= \frac{1}{16\pi^{3}}
\left[
\left|\psi^{\pm (u)}_{+}(x, \mathbf{k}_{\perp}^2)\right|^{2}
+
\left|\psi^{\pm (u)}_{-}(x, \mathbf{k}_{\perp}^2)\right|^{2}
\right].
\label{eq:scalarpartdesnityoperator}
\end{equation}

Similarly, for the vector-diquark sector, using Eq.~\eqref{eq:Vector_part fock state}, the density matrix can be expressed in terms of overlapped representation of LFWFs as

\begin{equation}
\begin{aligned}
\mathcal{P}^{A(\beta)}(x,\mathbf{k}_{\perp}^2; \pm)
&= \frac{1}{16\pi^3}\bigg[
\left|\psi^{\pm (\beta)}_{++}(x, \mathbf{k}_{\perp}^2)\right|^{2}
+\left|\psi^{\pm (\beta)}_{-+}(x, \mathbf{k}_{\perp}^2)\right|^{2}
+\left|\psi^{\pm (\beta)}_{+0}(x, \mathbf{k}_{\perp}^2)|\right|^{2} \\[4pt]
&\qquad\quad
+\left|\psi^{\pm (\beta)}_{-0}(x, \mathbf{k}_{\perp}^2)\right|^{2}
+\left|\psi^{\pm (\beta)}_{+-}(x, \mathbf{k}_{\perp}^2)\right|^{2}
+\left|\psi^{\pm (\beta)}_{--}(x, \mathbf{k}_{\perp}^2)\right|^{2}
\bigg].
\end{aligned}
\label{eq:vectorpartdesnityoperator}
\end{equation}

Using the light-front wave functions given in Eqs.~\eqref{eq:Scalar part wavefunction}, \eqref{eq: vector wavefunction for up}, and \eqref{eq: vector wavefunction for down}, we can write the density operators for the scalar-diquark sector and the vector-diquark sector for both spin-up and spin-down protons as
\begin{equation}
\mathcal{P}^{S(\beta)}(x,\mathbf{k}_{\perp}^2;\pm)
= N_{S}^{2}\,\frac{\log(1/x)}{\kappa^2\pi}
\left[
T_{1}^{(\beta)}(x)
+ \frac{\mathbf{k}_{\perp}^2}{M^{2}}\,T_{2}^{(\beta)}(x)
\right]
\exp\!\left[-R^{(\beta)}(x)\,\mathbf{k}_{\perp}^2\right].
\label{eq: densityopertor for scalar LFWF}
\end{equation}
Similarly, for the vector-diquark sector, one obtains
\begin{equation}
\begin{aligned}
\mathcal{P}^{A(\beta)}(x,\mathbf{k}_{\perp}^2;\pm)
&=\left(\frac{1}{3}N_0^{(\beta)2}+\frac{2}{3}N_1^{(\beta)2}\right)\frac{\log(1/x)}{\kappa^2\pi}\, 
\Bigg[
T_1^{(\beta)}(x)
+ \frac{\mathbf{k}_{\perp}^2}{M^2} T_2^{(\beta)}(x)
\Bigg]
\exp\!\left[-R^{(\beta)}(x)\,\mathbf{k}_{\perp}^2\right].
\end{aligned}
\label{eq: densityopertor for vector LFWF}
\end{equation}

where, we employ the following parameterizations\cite{Maji:2017bcz}

\begin{equation}
R^{(\beta)}(x) = \frac{\delta^{\beta}\,\log(1/x)}{\kappa^{2}(1-x)^{2}},
\end{equation}
\begin{equation}
T_{1}^{(\beta)}(x) = x^{2a_{1}^{\beta}}(1-x)^{2b_{1}^{\beta}-1},
\end{equation}
\begin{equation}
T_{2}^{(\beta)}(x) = x^{2a_{1}^{\beta}-2}(1-x)^{2b_{1}^{\beta}-1}.
\end{equation}

Using Eq.~\eqref{eg:main_equation} for different components of the TMD energy-momentum tensor, and substituting Eqs.~\eqref{eq: densityopertor for scalar LFWF} and \eqref{eq: densityopertor for vector LFWF} into Eqs.~\eqref{eg:a1}–\eqref{eg:a8}, we obtain the explicit expressions for the unpolarized gravitational TMDs as

\begin{equation}
\begin{aligned}
a_{1}^{(\beta)}(x,\mathbf{k}_{\perp}^2)
&=\mathcal{N}^{(\beta)}
\frac{x\,\log(1/x)}{\kappa^2 \pi}
\Bigg[
T_1^{(\beta)}(x)
+ \frac{\mathbf{k}_{\perp}^2}{M^2} T_2^{(\beta)}(x)
\Bigg]
\exp\!\left[-R^{(\beta)}(x)\, \mathbf{k}_{\perp}^2\right],
\end{aligned}
\label{eq:a1}
\end{equation}

\begin{equation}
\begin{aligned}
a_{3}^{(\beta)}(x,\mathbf{k}_{\perp}^2)
&=\mathcal{N}^{(\beta)}
\frac{\log(1/x)}{x\,\kappa^2 \pi}
\Bigg[
T_1^{(\beta)}(x)
+ \frac{\mathbf{k}_{\perp}^2}{M^2} T_2^{(\beta)}(x)
\Bigg]
\exp\!\left[-R^{(\beta)}(x)\, \mathbf{k}_{\perp}^2\right],
\end{aligned}
\label{eq:a3}
\end{equation}

\begin{equation}
\begin{aligned}
a_{5}^{(\beta)}(x,\mathbf{k}_{\perp}^2)
=\mathcal{N}^{(\beta)}\frac{\big(k_{\perp}^2+m_{q}^2\big)}{x M^2}
\frac{\log(1/x)}{\kappa^2 \pi}
\Bigg[
T_1^{(\beta)}(x)
+ \frac{\mathbf{k}_{\perp}^2}{M^2} T_2^{(\beta)}(x)
\Bigg]
\exp\!\left[-R^{(\beta)}(x)\, \mathbf{k}_{\perp}^2\right]- a_{1}^{(\beta)}(x,\mathbf{k}_{\perp}^2),
\end{aligned}\
\label{eq:a5}
\end{equation}

\begin{equation}
\begin{aligned}
a_{6}^{(\beta)}(x,\mathbf{k}_{\perp}^2)
&=
\mathcal{N}^{(\beta)}
\frac{\log(1/x)}{\kappa^2 \pi}
\Bigg[
T_1^{(\beta)}(x)
+ \frac{\mathbf{k}_{\perp}^2}{M^2} T_2^{(\beta)}(x)
\Bigg]
\exp\!\left[-R^{(\beta)}(x)\, \mathbf{k}_{\perp}^2\right],
\end{aligned}
\label{eq:a6}
\end{equation}

\begin{equation}
\begin{aligned}
a_{7}^{(\beta)}(x,\mathbf{k}_{\perp}^2)
&=
\mathcal{N}^{(\beta)}
\frac{\log(1/x)}{\kappa^2 \pi}
\Bigg[
T_1^{(\beta)}(x)
+ \frac{\mathbf{k}_{\perp}^2}{M^2} T_2^{(\beta)}(x)
\Bigg]
\exp\!\left[-R^{(\beta)}(x)\, \mathbf{k}_{\perp}^2\right],
\end{aligned}
\label{eq:a7}
\end{equation}

\begin{equation}
\begin{aligned}
a_{8}^{(\beta)}(x,\mathbf{k}_{\perp}^2)
=\mathcal{N}^{(\beta)}\frac{\big(\mathbf{k}_{\perp}^2+m_{q}^2\big)}{x^2 M^2}
\frac{\log(1/x)}{\kappa^2 \pi}
\Bigg[
T_1^{(\beta)}(x)
+ \frac{\mathbf{k}_{\perp}^2}{M^2} T_2^{(\beta)}(x)
\Bigg]\exp\!\left[-R^{(\beta)}(x)\, \mathbf{k}_{\perp}^2\right]- a_{6}^{(\beta)}(x,\mathbf{k}_{\perp}^2).
\end{aligned}
\label{eq:a8}
\end{equation}

The normalization factor $\mathcal{N}^{(\beta)}$ appearing in Eqs.~\eqref{eq:a1}--\eqref{eq:a8} is given by
\begin{equation}
\mathcal{N}^{(\beta)}=
\Bigg(
N_s^{(\beta)2}\, C_S^2
+ C_A^2
\left(
\frac{1}{3}N_0^{(\beta)2}
+ \frac{2}{3}N_1^{(\beta)2}
\right)
\Bigg).
\end{equation}
 
where $C_A=C_V, C_{VV}$ for up and down quark respectively.

\vspace{0.5em}

Thus, Eqs.~\eqref{eq:a1}--\eqref{eq:a8}, we obtain the analytical expressions for the unpolarized (T-even)      gravitational TMDs $a_{i}^{(\beta)}(x,k_\perp^2)$  within the light-front quark--diquark model (LFQDM). These gravitational TMDs are related to the mechanical properties of the proton.  In the next section, we discuss their behavior in  three-dimensions and interpret them in terms of the proton's mechanical structure. We further verify that the gravitational TMDs obtained above, given in Eqs.~\eqref{eq:a1}–\eqref{eq:a8}, satisfy the model-independent relations derived in Ref.~\cite{Lorce:2023zzg}. These relations establish a connection between the quark gravitational TMDs and the twist-2, twist-3, and twist-4 unpolarized quark TMDs and therefore provide a nontrivial and strong consistency check of our results. The corresponding unpolarized quark TMDs within the light-front quark-diquark model have been studied extensively in the literature; see Refs.~\cite{Maji:2017bcz, Sharma:2023wha}. The corresponding model-independent relations are given by
\begin{equation}
\begin{aligned}
a_{1}^{(\beta)}(x,\mathbf{k}_{\perp}^2) &= x f_{1}^{(\beta)}(x,\mathbf{k}_{\perp}^2),
\end{aligned}
\label{eq:model-independent-relation1}
\end{equation}
\begin{equation}
\begin{aligned}
a_{7}^{(\beta)}(x,\mathbf{k}_{\perp}^2) &= x f^{\perp (\beta)}(x,\mathbf{k}_{\perp}^2),
\end{aligned}
\label{eq:model-independent-relation2}
\end{equation}
\begin{equation}
\begin{aligned}
a_{3}^{(\beta)}(x,\mathbf{k}_{\perp}^2) &= f^{\perp (\beta)}(x,\mathbf{k}_{\perp}^2),
\end{aligned}
\label{eq:model-independent-relation3}
\end{equation}
\begin{equation}
\begin{aligned}
a_{1}^{(\beta)}(x,\mathbf{k}_{\perp}^2) + a_{5}^{(\beta)}(x,\mathbf{k}_{\perp}^2) &= x f_{3}^{(\beta)}(x,\mathbf{k}_{\perp}^2),
\end{aligned}
\label{eq:model-independent-relation4}
\end{equation}
\begin{equation}
\begin{aligned}
a_{6}^{(\beta)}(x,\mathbf{k}_{\perp}^2) &= f_{1}^{(\beta)}(x,\mathbf{k}_{\perp}^2),
\end{aligned}
\label{eq:model-independent-relation5}
\end{equation}
\begin{equation}
\begin{aligned}
a_{6}^{(\beta)}(x,\mathbf{k}_{\perp}^2) + a_{8}^{(\beta)}(x,\mathbf{k}_{\perp}^2) &= f_{3}^{(\beta)}(x,\mathbf{k}_{\perp}^2).
\end{aligned}
\label{eq:model-independent-relation6}
\end{equation}

In addition to the relations discussed above, if we assume a symmetric transverse-momentum-dependent energy-momentum tensor in Eq.~\eqref{eg: parameterizedTMD}, we find that the components $\mathcal{T}_{q}^{+i}$ and $\mathcal{T}_{q}^{i+}$ are equal, which implies
\begin{equation}
\mathcal{T}_{q}^{+i} = \mathcal{T}_{q}^{i+}
\;\;\Longrightarrow\;\;
a_{6}^{(\beta)}(x,\mathbf{k}_{\perp}^2)
=
a_{7}^{(\beta)}(x, \mathbf{k}_{\perp}^2),
\label{eq:modeld-dependent-relation}
\end{equation}
This result is consistent with our explicit model calculations presented above in Eqs.~\eqref{eq:a6} and \eqref{eq:a7}. Furthermore, by substituting Eqs.~\eqref{eq:model-independent-relation2} and \eqref{eq:model-independent-relation5} into Eq.~\eqref{eq:modeld-dependent-relation}, we obtain the relation
\begin{equation}
f_{1}^{(\beta)}(x,\mathbf{k}_{\perp}^2)
=
x\, f^{\perp(\beta)}(x,\mathbf{k}_{\perp}^2).
\label{eq:freequarkmotion}
\end{equation}
The relation in Eq.~\eqref{eq:freequarkmotion} is also found in the light-front quark-diquark model (LFQDM)~\cite{Maji:2017bcz, Sharma:2023wha} and in the light-front constituent quark model (LFCQM)~\cite{Lorce:2014hxa}. In full QCD, such relations are not expected to hold in general due to the presence of quark-gluon interactions. However, both the LFQDM and the LFCQM are effective models in which explicit gluonic degrees of freedom are absent, and such relations naturally emerge.

\subsection{Gravitational PDFs}

In this subsection, we compute the unpolarized quark gravitational parton distribution functions (PDFs). Since the gravitational TMDs $a_i^{(\beta)}(x,k_\perp^2)$ depend on both the longitudinal momentum fraction $x$ and the transverse momentum $\boldsymbol{k}_\perp^2$, the associated gravitational PDFs are obtained by integrating over the transverse momentum,i.e.,

\begin{equation}
A_{i}^{(\beta)}(x) = \int d^2\mathbf{k}_\perp\, a_{i}^{(\beta)}(x,k_\perp^2).
\end{equation}

Substituting Eqs.~\eqref{eq:a1}--\eqref{eq:a8} into the above relation, we obtain the corresponding gravitational PDFs as

\begin{equation}
A_{1}^{(\beta)}(x)
=
\mathcal{N}^{(\beta)}
\frac{x\,\log(1/x)}{\kappa^{2}}
\left[
\frac{T_{1}^{(\beta)}(x)}{R^{(\beta)}(x)}
+
\frac{T_{2}^{(\beta)}(x)}{M^{2} R^{(\beta)}(x)^{2}}
\right],
\end{equation}

\begin{equation}
A_{3}^{(\beta)}(x)
=\mathcal{N}^{(\beta)}
\frac{\log(1/x)}{\kappa^{2}}
\left[
\frac{T_{1}^{(\beta)}(x)}{x\,R^{(\beta)}(x)}
+
\frac{T_{2}^{(\beta)}(x)}{x\,M^{2} R^{(\beta)}(x)^{2}}
\right],
\end{equation}

\begin{equation}
\begin{aligned}
A_{5}^{(\beta)}(x)
&=\mathcal{N}^{(\beta)}
\frac{\log(1/x)}{\kappa^{2}}
\Bigg\{
\frac{1}{x M^{2}}
\left[
\frac{T_{1}^{(\beta)}(x)}{R^{(\beta)}(x)^{2}}
+
\frac{2 T_{2}^{(\beta)}(x)}{M^{2}R^{(\beta)}(x)^{3}}
+
\frac{m_q^{2} T_{1}^{(\beta)}(x)}{R^{(\beta)}(x)}
+
\frac{m_q^{2} T_{2}^{(\beta)}(x)}{M^{2} R^{(\beta)}(x)^{2}}
\right]
\Bigg\}-A^{(\beta)}_{1}(x).
\end{aligned}
\label{eq:PDF5}
\end{equation}

The above expressions define the unpolarized quark gravitational PDFs $A_{i}^{(\beta)}(x)$ obtained from the corresponding gravitational TMDs in Eqs.~\eqref{eq:a1}–\eqref{eq:a5}. Only these three gravitational PDFs are non-zero, whereas the remaining unpolarized gravitational PDFs associated with the gravitational TMDs in Eqs.~\eqref{eq:a6}–\eqref{eq:a8} are zero, this follows from the parametrization in Eq.~\eqref{eg: parameterizedTMD}, where the corresponding terms appear with factors linear in $\mathbf{k}_\perp$, which necessarily give zero upon performing the integration $\int d^2\mathbf{k}_\perp\, \mathcal{T}^{\mu\nu}$. As an additional consistency check, the resulting gravitational PDFs are found to satisfy the similar model-independent relations as those presented in Eqs.~\eqref{eq:model-independent-relation1}–\eqref{eq:model-independent-relation6}.

\section{Numerical result and Discussion}
\label{sec:Numerical result and Discussion}
In this section, we present the numerical results for the gravitational TMDs and gravitational PDFs calculated within the light-front quark-diquark model described above. The numerical values of the model parameters used in the plots are listed in Table~\ref{tab:Normalizationconstant} and are taken at the initial model scale $\mu_{0}=0.313~\mathrm{GeV}$.

\subsection{Unpolarized gravitational-TMDs}


\begin{figure}[htbp]
\centering

\includegraphics[width=0.32\linewidth]{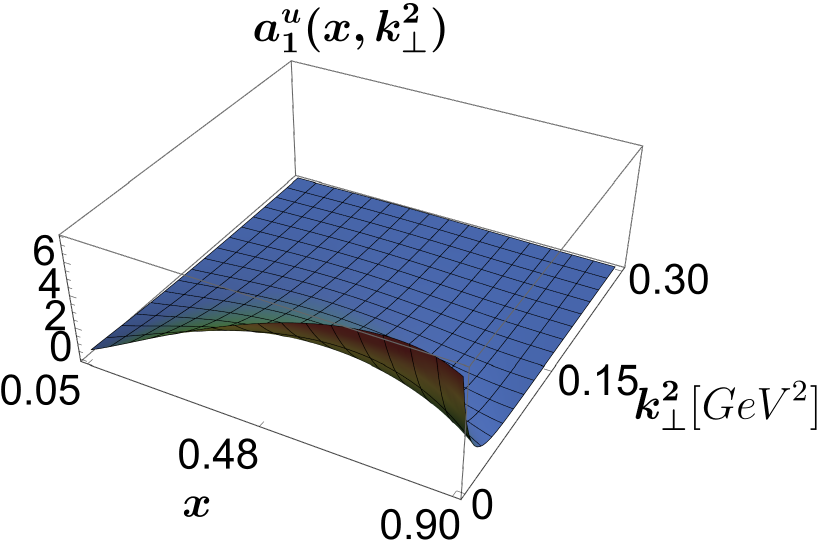}
\hfill
\includegraphics[width=0.32\linewidth]{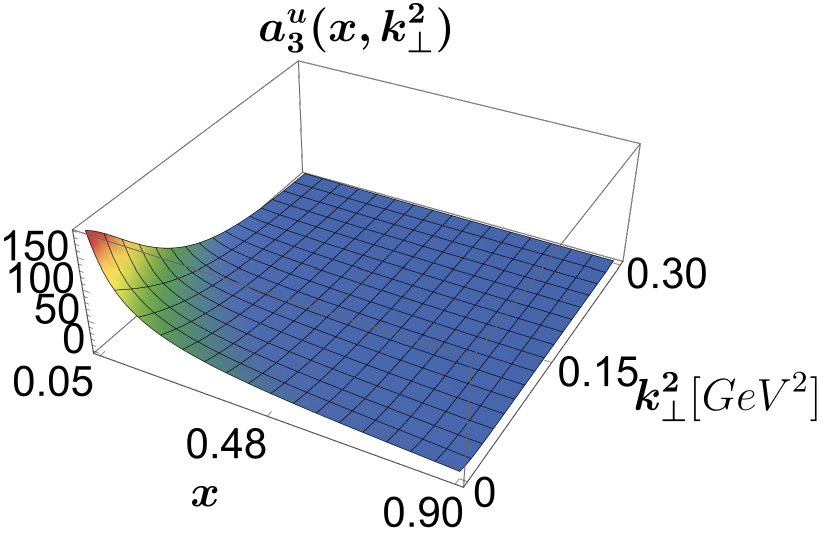}
\hfill
\includegraphics[width=0.32\linewidth]{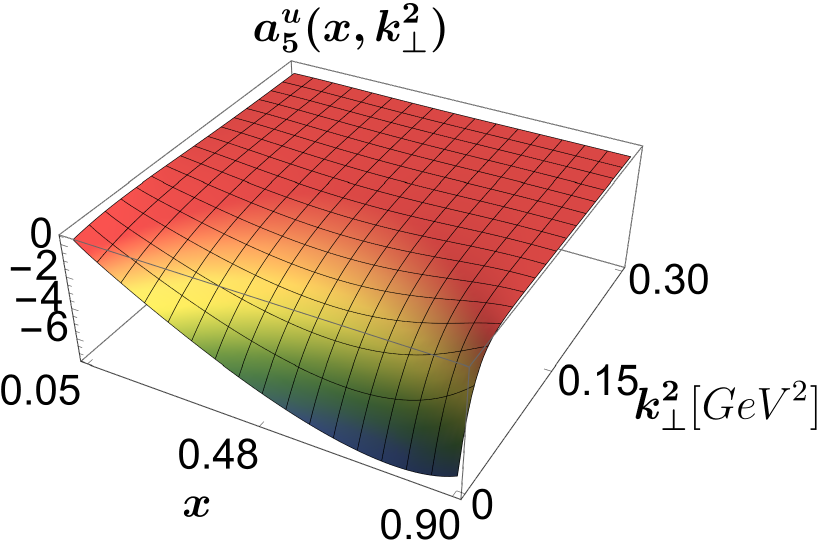}

\vspace{0.3cm}

\includegraphics[width=0.32\linewidth]{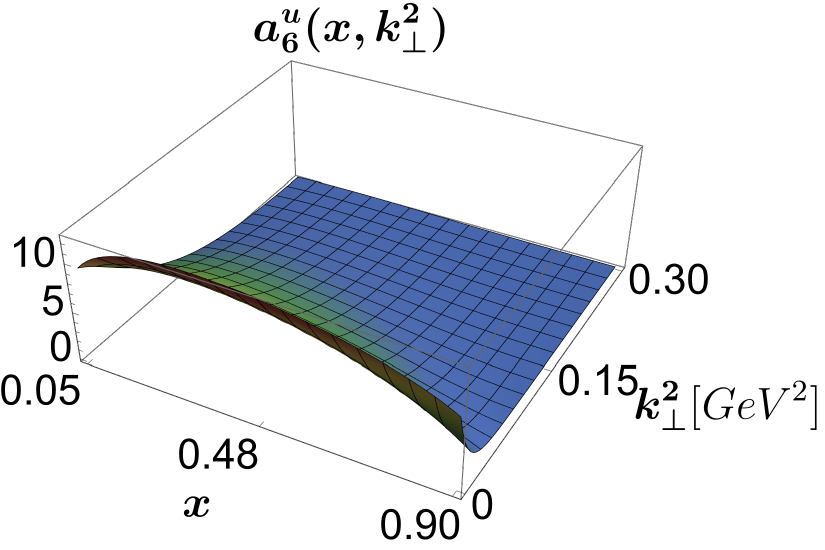}
\hfill
\includegraphics[width=0.32\linewidth]{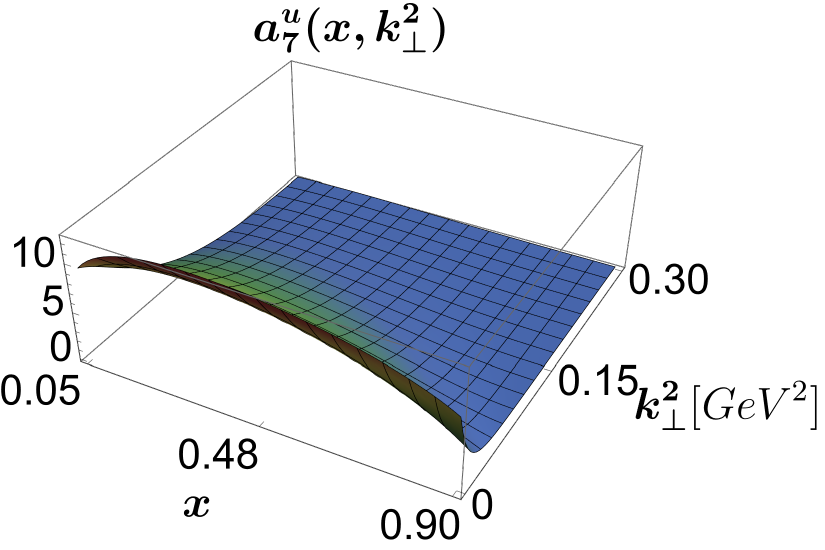}
\hfill
\includegraphics[width=0.32\linewidth]{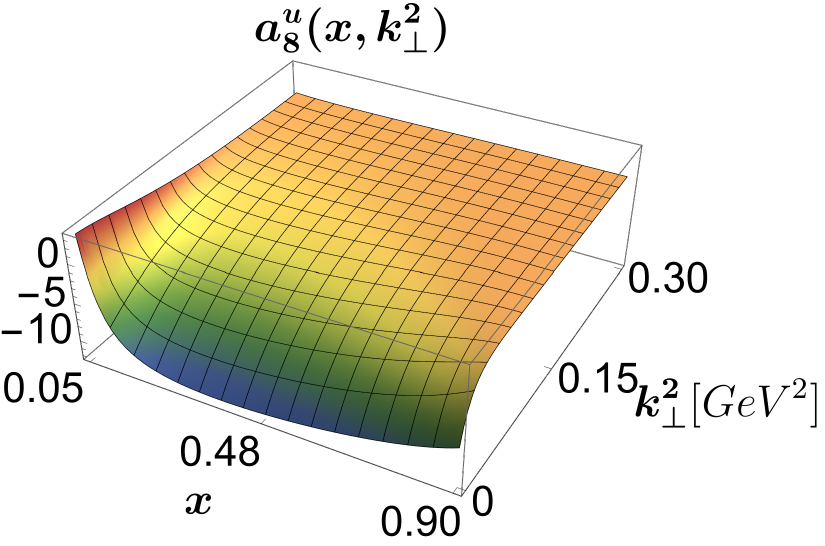}

\caption{
The gravitational transverse momentum distributions 
$a_{1}^{(\beta)}(x,\mathbf{k}_{\perp}^{2})$, 
$a_{3}^{(\beta)}(x,\mathbf{k}_{\perp}^{2})$, 
$a_{5}^{(\beta)}(x,\mathbf{k}_{\perp}^{2})$, 
$a_{6}^{(\beta)}(x,\mathbf{k}_{\perp}^{2})$, 
$a_{7}^{(\beta)}(x,\mathbf{k}_{\perp}^{2})$, and 
$a_{8}^{(\beta)}(x,\mathbf{k}_{\perp}^{2})$ for the up quark, shown as functions of $x$ and $\mathbf{k}_\perp^2$ at the initial scale $\mu_0$.
}
\label{fig:up}
\end{figure}

\begin{figure}[htbp]
\centering

\includegraphics[width=0.32\linewidth]{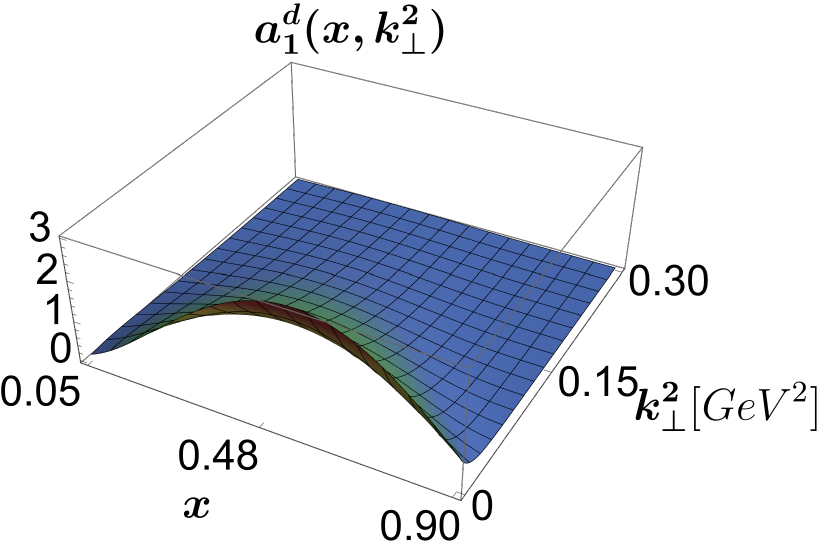}
\hfill
\includegraphics[width=0.32\linewidth]{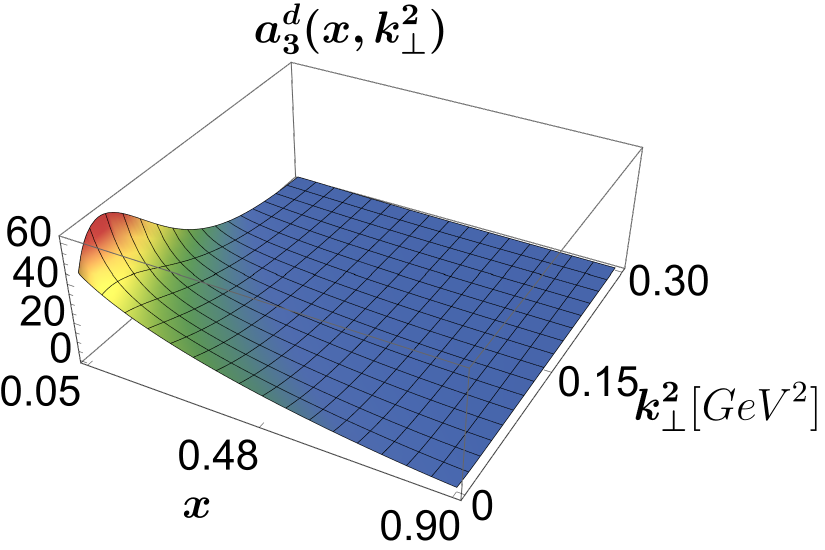}
\hfill
\includegraphics[width=0.32\linewidth]{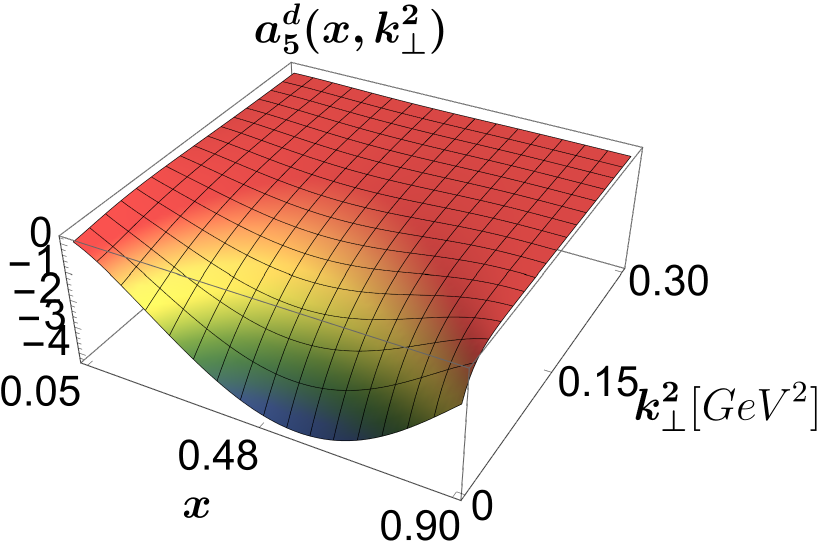}

\vspace{0.3cm}

\includegraphics[width=0.32\linewidth]{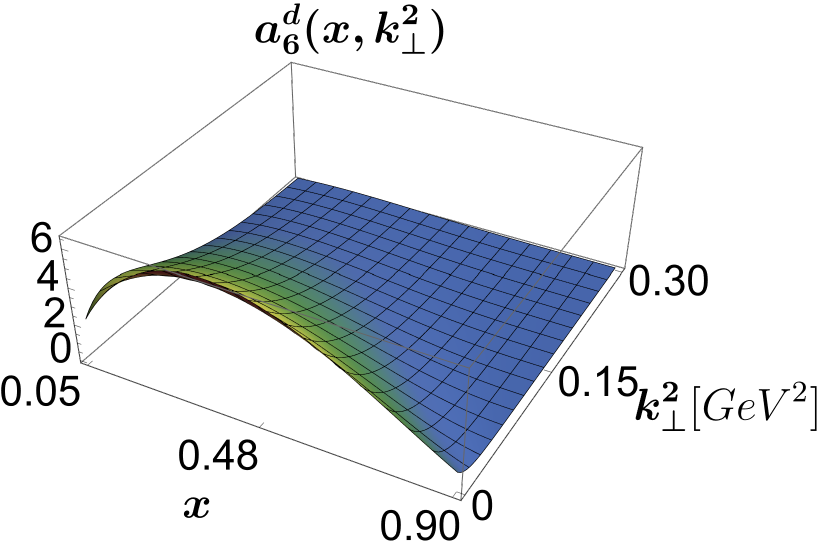}
\hfill
\includegraphics[width=0.32\linewidth]{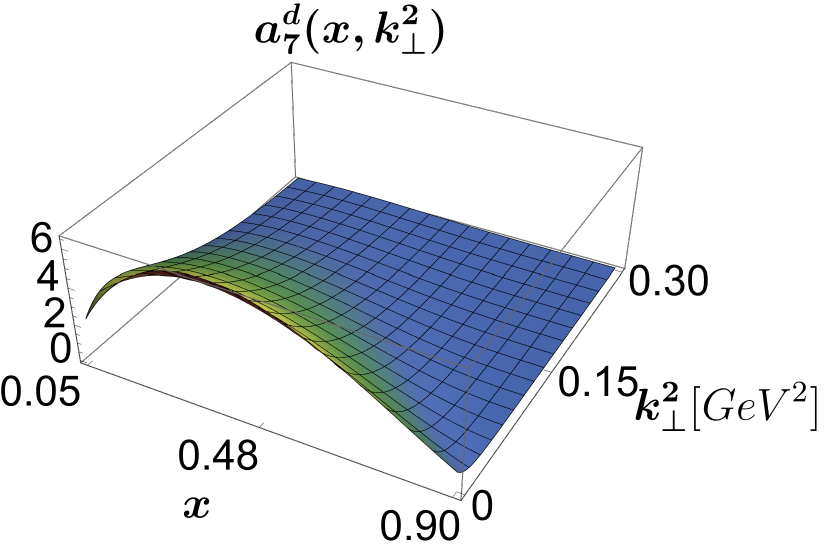}
\hfill
\includegraphics[width=0.32\linewidth]{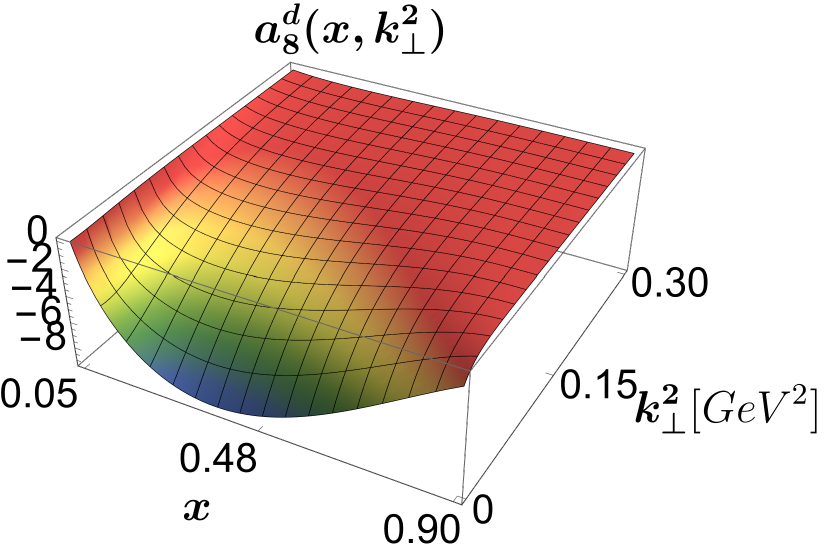}

\caption{The gravitational transverse momentum distributions 
$a_{1}^{(\beta)}(x,\mathbf{k}_{\perp}^{2})$, 
$a_{3}^{(\beta)}(x,\mathbf{k}_{\perp}^{2})$, 
$a_{5}^{(\beta)}(x,\mathbf{k}_{\perp}^{2})$, 
$a_{6}^{(\beta)}(x,\mathbf{k}_{\perp}^{2})$, 
$a_{7}^{(\beta)}(x,\mathbf{k}_{\perp}^{2})$, and 
$a_{8}^{(\beta)}(x,\mathbf{k}_{\perp}^{2})$ for the down quark, shown as functions of $x$ and $\mathbf{k}_\perp^2$ at the initial scale $\mu_0$.
}
\label{fig:down}
\end{figure}

 
\begin{figure}[htbp]
    \centering
    \includegraphics[width=0.37\linewidth]{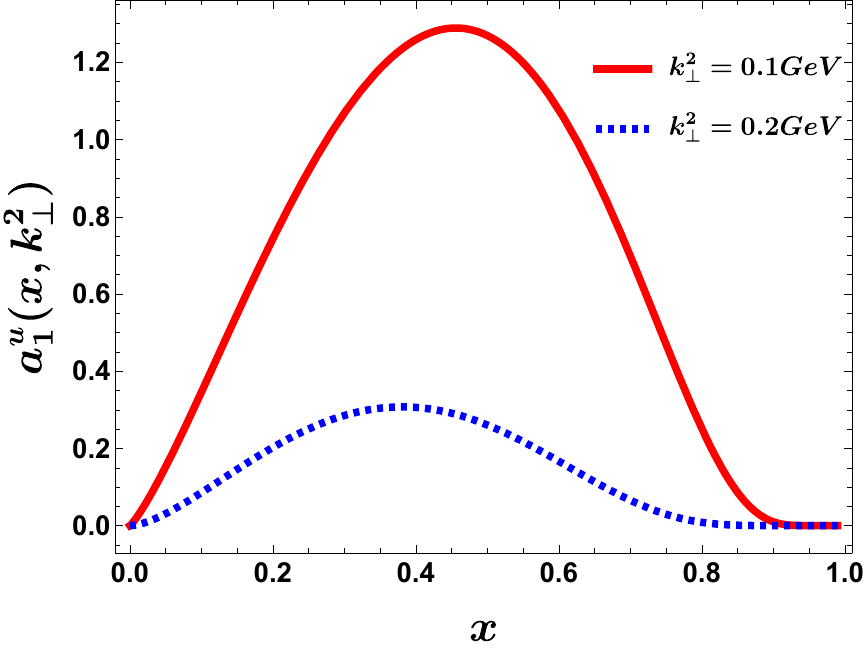}
    \hspace{0.04\linewidth}
    \includegraphics[width=0.37\linewidth]{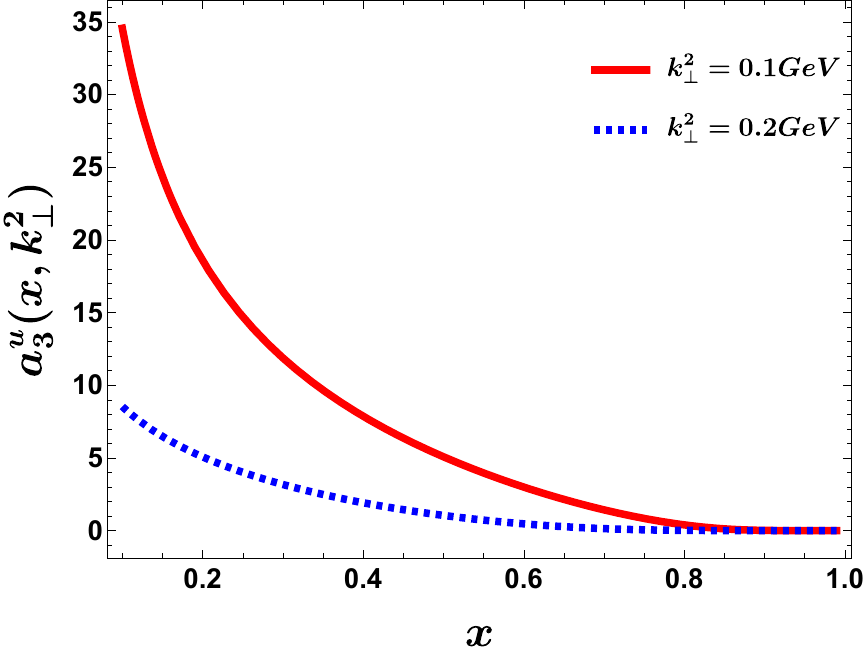}
    \includegraphics[width=0.37\linewidth]{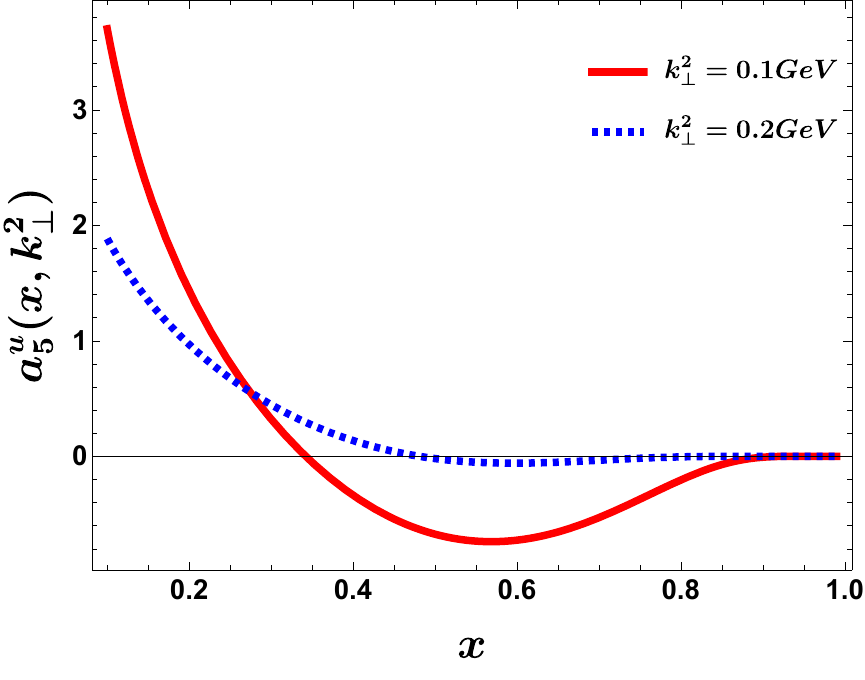}
    \hspace{0.04\linewidth}
    \includegraphics[width=0.37\linewidth]{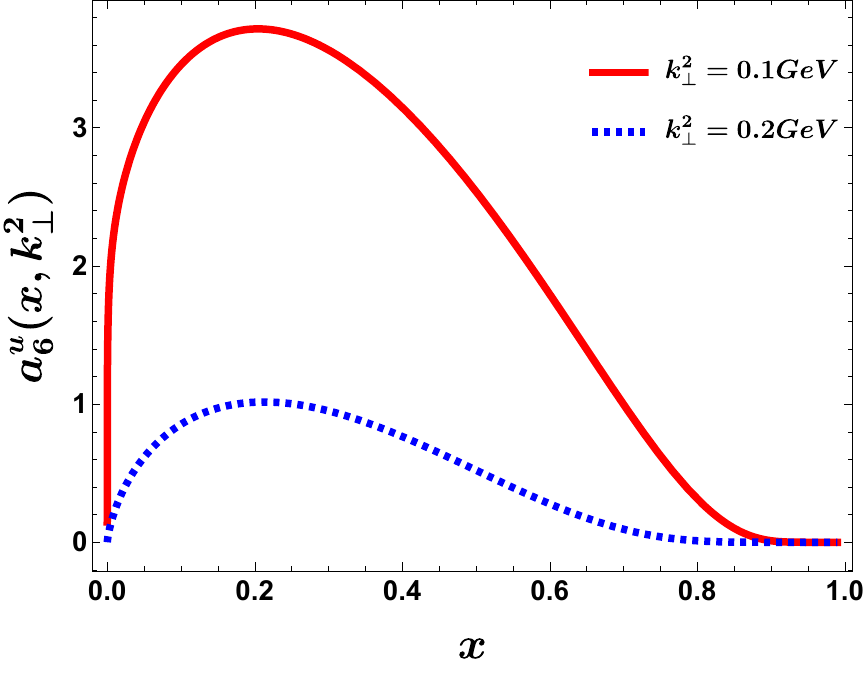}\\

    \begin{center}
        \includegraphics[width=0.37\linewidth]{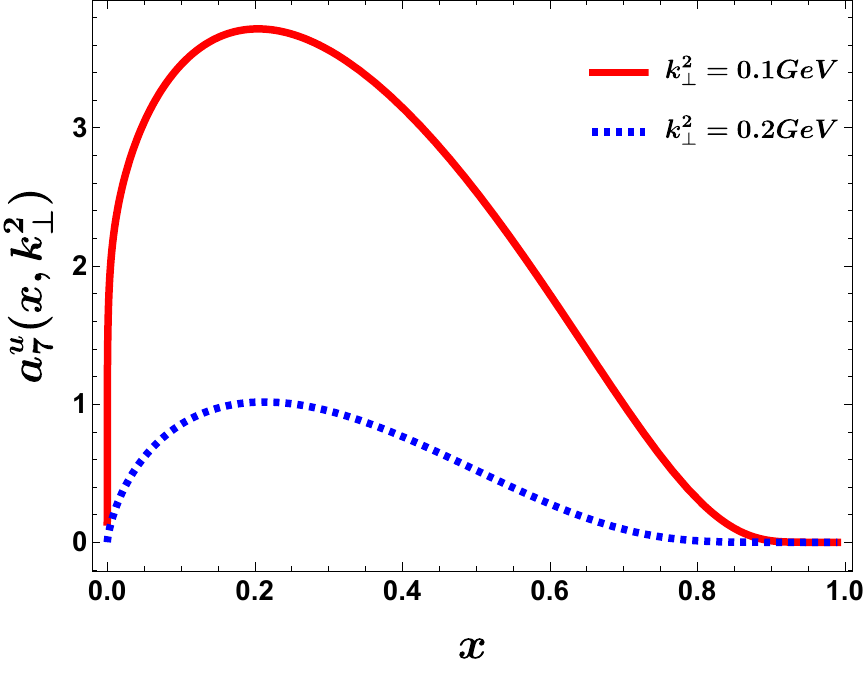}
        \hspace{0.04\linewidth}
    \includegraphics[width=0.37\linewidth]{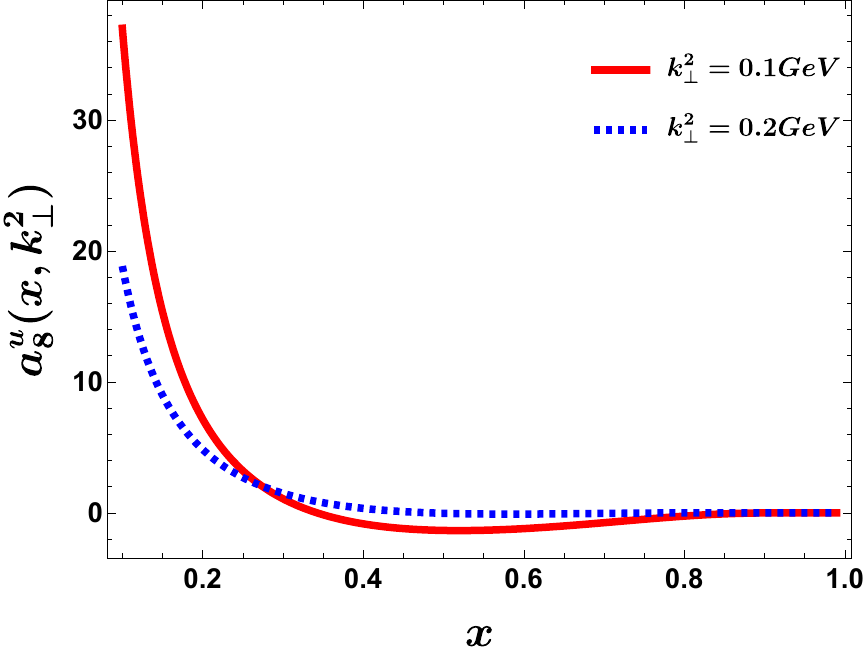}
    \end{center}
    \caption{
 The unpolarized gravitational transverse-momentum distributions 
$a_{1}^{(\beta)}(x,\mathbf{k}_{\perp}^{2})$, 
$a_{3}^{(\beta)}(x,\mathbf{k}_{\perp}^{2})$, 
$a_{5}^{(\beta)}(x,\mathbf{k}_{\perp}^{2})$, 
$a_{6}^{(\beta)}(x,\mathbf{k}_{\perp}^{2})$, 
$a_{7}^{(\beta)}(x,\mathbf{k}_{\perp}^{2})$, and 
$a_{8}^{(\beta)}(x,\mathbf{k}_{\perp}^{2})$ for the up quark, shown as functions of $x$ for different values of $\mathbf{k}_{\perp}^{2}$.
    }
    \label{fig:upx}
\end{figure}

\begin{figure}[htbp]
    \centering
    \includegraphics[width=0.37\linewidth]{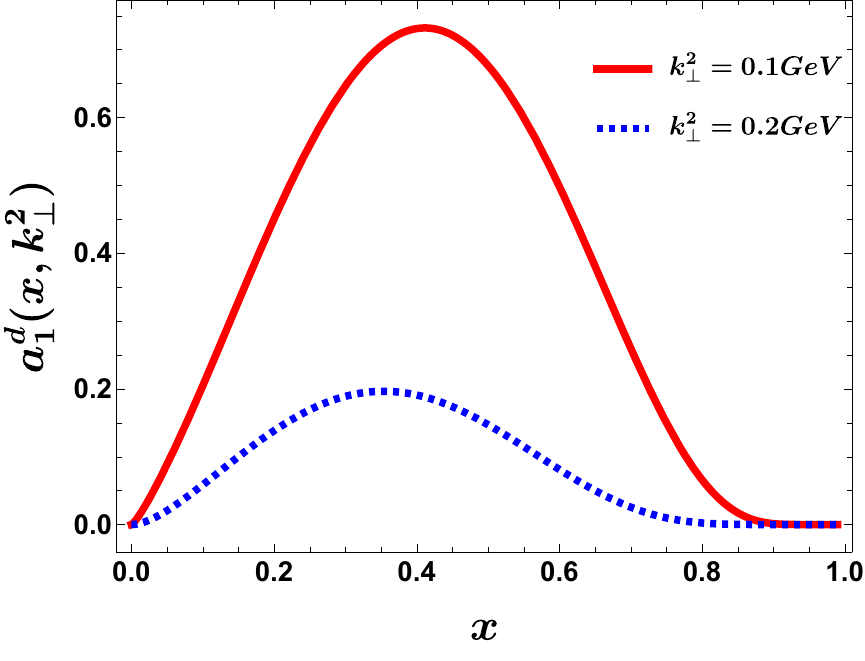}
    \hspace{0.04\linewidth}
    \includegraphics[width=0.37\linewidth]{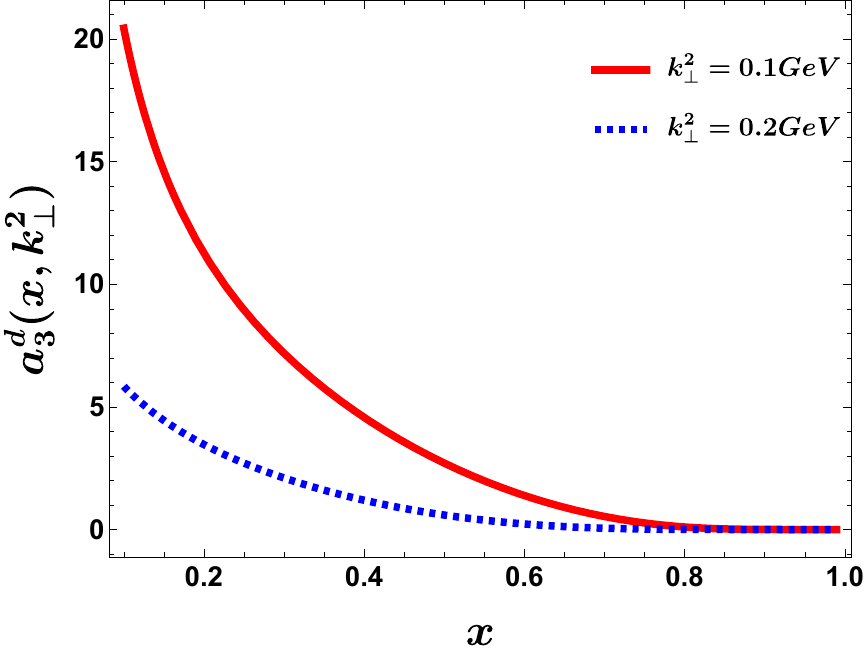}
    
    \includegraphics[width=0.37\linewidth]{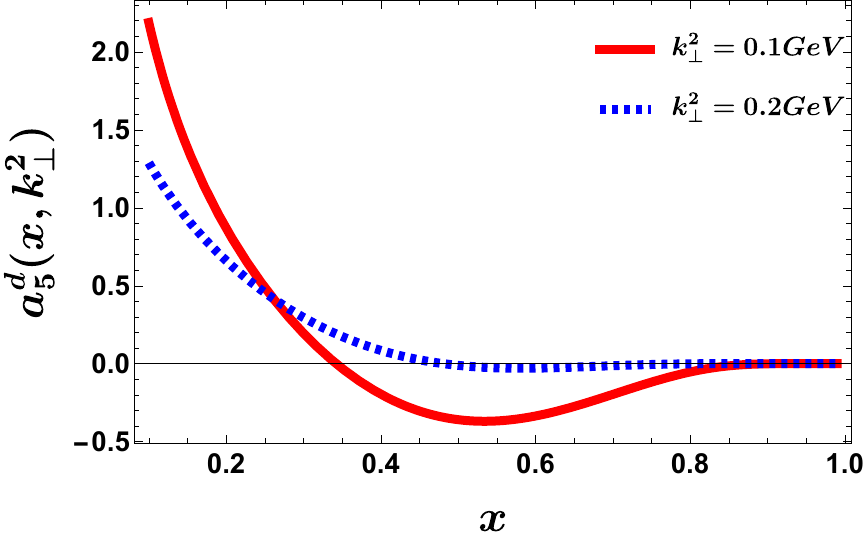}
    \hspace{0.04\linewidth}
    \includegraphics[width=0.37\linewidth]{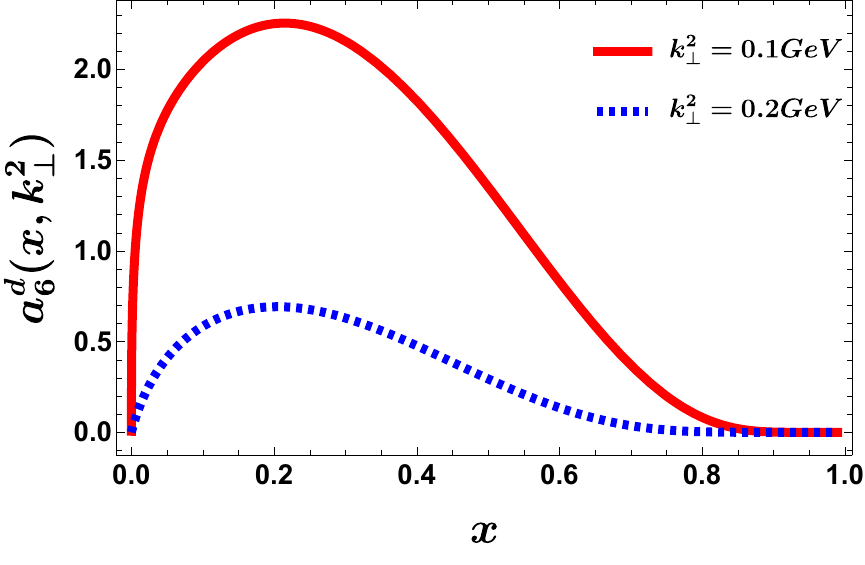}\\
    \includegraphics[width=0.37\linewidth]{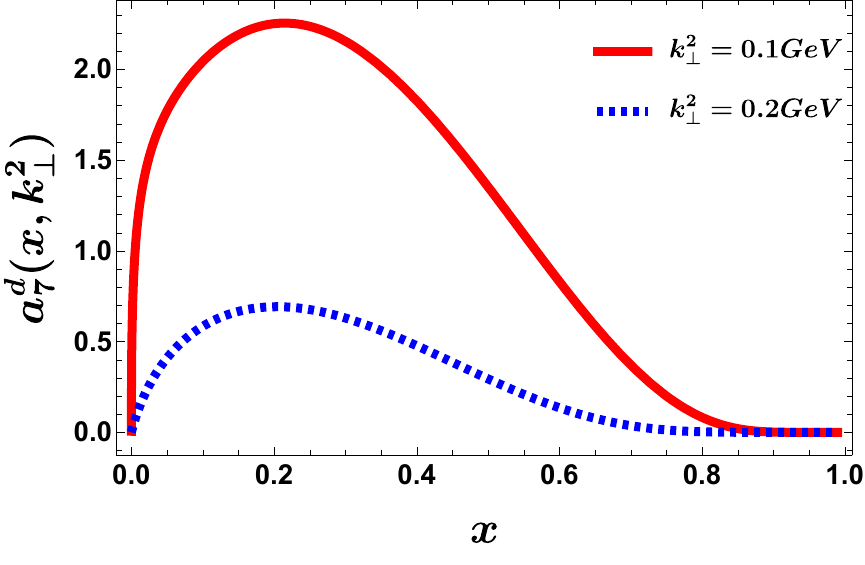}
    \hspace{0.04\linewidth}
    \includegraphics[width=0.37\linewidth]{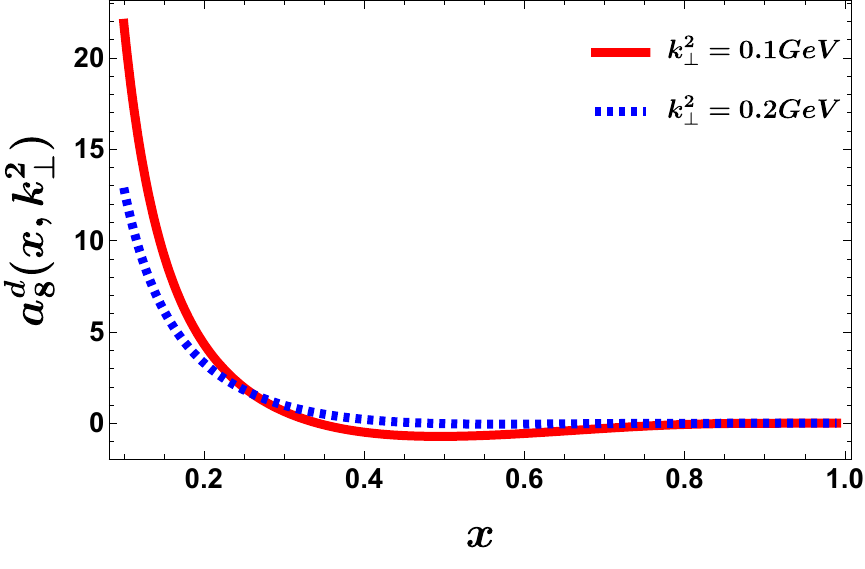}
    \caption{The unpolarized gravitational transverse-momentum distributions 
$a_{1}^{(\beta)}(x,\mathbf{k}_{\perp}^{2})$, 
$a_{3}^{(\beta)}(x,\mathbf{k}_{\perp}^{2})$, 
$a_{5}^{(\beta)}(x,\mathbf{k}_{\perp}^{2})$, 
$a_{6}^{(\beta)}(x,\mathbf{k}_{\perp}^{2})$, 
$a_{7}^{(\beta)}(x,\mathbf{k}_{\perp}^{2})$, and 
$a_{8}^{(\beta)}(x,\mathbf{k}_{\perp}^{2})$ for the down quark, shown as functions of $x$ for different values of $\mathbf{k}_{\perp}^{2}$.}
    \label{fig:downx}
\end{figure}

We present the three-dimensional behavior of the gravitational transverse-momentum-dependent distributions as functions of $x$ and $\mathbf{k}_\perp^2$ within the light-front quark-diquark (LFQD) model in Figures.~\ref{fig:up} and \ref{fig:down} for the up and down quarks, respectively. The distributions $a_{1}^{(\beta)}(x,\mathbf{k}_\perp^2)$, $a_{3}^{(\beta)}(x,\mathbf{k}_\perp^2)$, $a_{6}^{(\beta)}(x,\mathbf{k}_\perp^2)$, and $a_{7}^{(\beta)}(x,\mathbf{k}_\perp^2)$ exhibit positive peaks for both $u$ and $d$ quarks. This behavior is consistent with expectations, as these gravitational TMDs are related to the unpolarized twist-2 and twist-3 TMDs, which are known to be positive within this model~\cite{Maji:2017bcz, Sharma:2023wha}. 

In contrast, the gravitational TMDs $a_{5}^{(\beta)}(x,\mathbf{k}_\perp^2)$ and $a_{8}^{(\beta)}(x,\mathbf{k}_\perp^2)$ exhibit a qualitatively different behavior, characterized by predominantly negative distributions for both quark flavors. This feature can be understood from Eqs.~\eqref{eq:a5} and \eqref{eq:a8}. In the small-$x$ region ($x \to 0$), the first terms in both $a_{5}^{(\beta)}$ and $a_{8}^{(\beta)}$ dominate and lead to a rapidly increasing (divergent-like) behavior. However, as $x$ increases, the relative contribution of the second terms becomes significant and eventually dominates, resulting in an overall negative contribution to the distributions. This behavior is clearly illustrated in Figs.~\ref{fig:upx} and \ref{fig:downx}, where both $a_{5}^{(\beta)}(x,\mathbf{k}_\perp^2)$ and $a_{8}^{(\beta)}(x,\mathbf{k}_\perp^2)$ are shown as functions of $x$ for different values of $\mathbf{k}_\perp^2$.

We plot the gravitational TMDs as functions of $x$ for different values of $\mathbf{k}_\perp^{2}$, namely $\mathbf{k}_\perp^{2}=0.1~\mathrm{GeV}^{2}$ (red curves) and $\mathbf{k}_\perp^{2}=0.2~\mathrm{GeV}^{2}$ (dashed blue curves), in Figs.~\ref{fig:upx} and \ref{fig:downx} for the up and down quarks, respectively. We observe that, as the transverse momentum $\mathbf{k}_\perp$ increases, the peak of the gravitational TMDs $a_i^{(\beta)}(x,\mathbf{k}_\perp^{2})$ decreases for both $u$ and $d$ quarks. Moreover, all gravitational TMDs decrease with increasing $x$ and vanish in the large-$x$ limit. This behavior originates from the soft-wall AdS/QCD--inspired light-front wave function in Eq.~\eqref{eq: Ads/ACD wavefunction}, which contains the Gaussian suppression factor
$\exp\!\left[-\mathbf{k}_\perp^{2}/\big(2\kappa^{2}x(1-x)\big)\right]$, leading to strong suppression near the kinematic endpoint $x \to 1$ and reflecting the confining dynamics of quarks in the model. 

In addition, for both $u$ and $d$ quarks, Figures.~\ref{fig:upx} and \ref{fig:downx} show that the gravitational TMDs $a_{1}^{(\beta)}(x,\mathbf{k}_\perp^{2})$, $a_{6}^{(\beta)}(x,\mathbf{k}_\perp^{2})$, and $a_{7}^{(\beta)}(x,\mathbf{k}_\perp^{2})$ are suppressed and approach zero as $x\to 0$. This behavior is again driven by the Gaussian exponential factor in the wave function in Eq.~\eqref{eq: Ads/ACD wavefunction}. By contrast, the gravitational TMDs $a_{3}^{(\beta)}(x,\mathbf{k}_\perp^{2})$, $a_{5}^{(\beta)}(x,\mathbf{k}_\perp^{2})$, and $a_{8}^{(\beta)}(x,\mathbf{k}_\perp^{2})$ exhibit a divergence as $x\to 0$, which arises from their explicit $1/x$ and $1/x^{2}$ dependence.

\subsection{Unpolarized gravitational-PDFs}

\begin{figure}[htbp]
    \centering
    \includegraphics[width=0.388\linewidth]{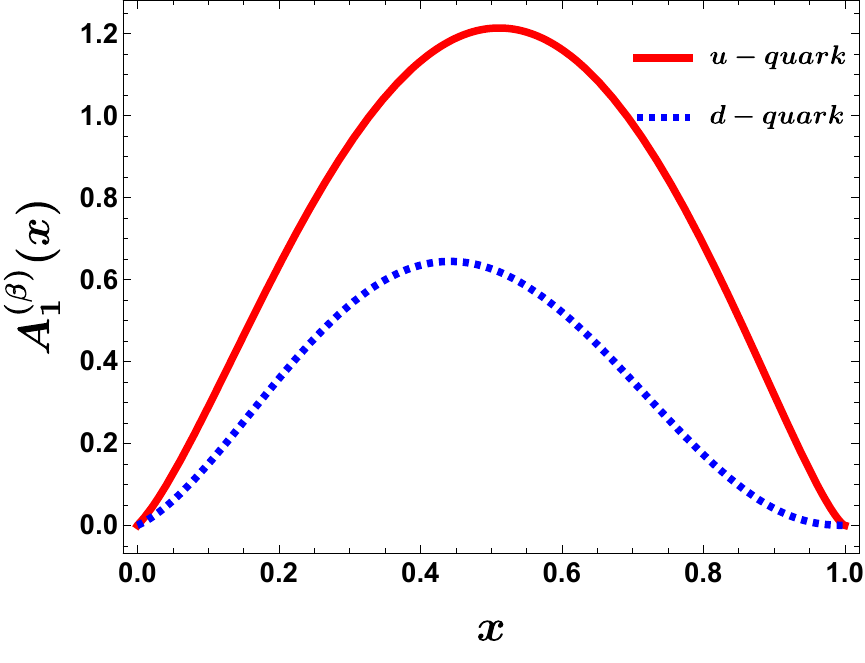}
    \hspace{0.04\linewidth}
    \includegraphics[width=0.388\linewidth]{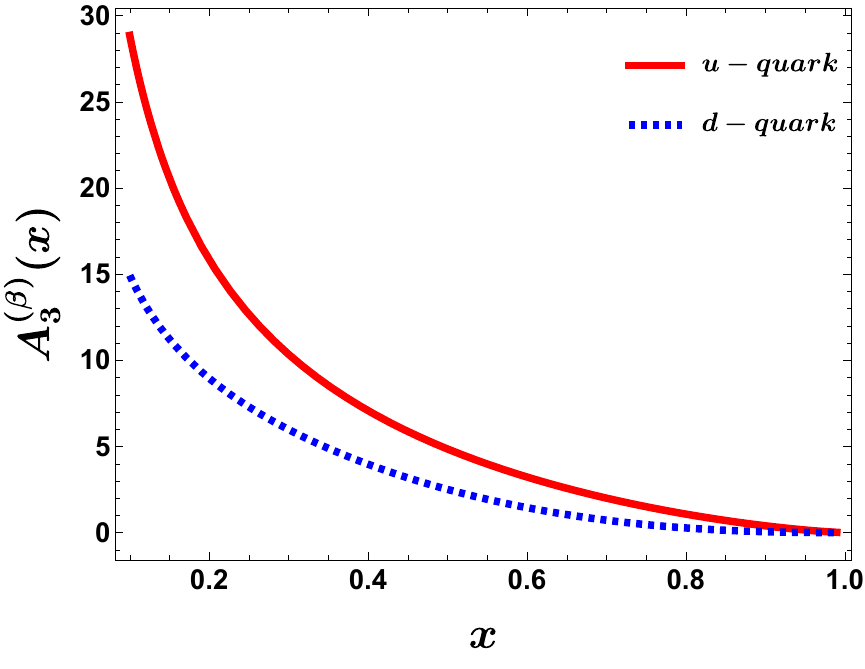}
    \hspace{0.02\linewidth}
    \includegraphics[width=0.388\linewidth]{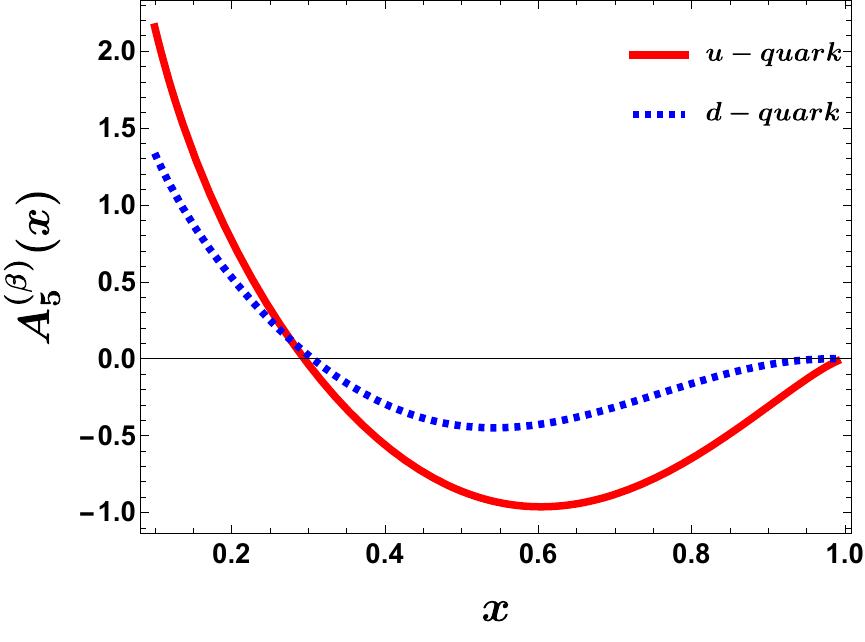}
    \caption{Unpolarized gravitational parton distribution functions $A_{1}^{(\beta)}(x)$, $A_{3}^{(\beta)}(x)$, and $A_{5}^{(\beta)}(x)$ for up and down quarks, shown as functions of the momentum fraction $x$.}
    \label{fig:PDFs}
\end{figure}

We present the unpolarized gravitational PDFs $A_{1}^{(\beta)}(x)$, $A_{3}^{(\beta)}(x)$, and $A_{5}^{(\beta)}(x)$ for both up and down quarks in Fig.~\ref{fig:PDFs}. The distribution $A_{1}^{(\beta)}(x)$ exhibits a valence-like structure for both flavors: it vanishes at the kinematic endpoints $x\to 0$ and $x\to 1$, and develops a broad maximum in the intermediate region $x\sim 0.4$--$0.5$, which is valence quark domain reflecting quark number density. The endpoint suppression is governed by the Gaussian ansatz of the soft-wall AdS/QCD wave function in Eq.~\eqref{eq: Ads/ACD wavefunction}, consistent with the behavior observed for the corresponding gravitational TMDs in Figs.~\ref{fig:upx} and \ref{fig:downx}. Quantitatively, the $u$-quark contribution is significantly larger than the $d$-quark contribution across the entire $x$ range. This hierarchy reflects the proton’s valence structure and follows from the flavor decomposition in Eq.~\eqref{eg:proton-state}.

For large $x$, both $A_{3}^{(\beta)}(x)$ and $A_{5}^{(\beta)}(x)$ approach zero as $x\to 1$ due to the Gaussian suppression inherited from the soft-wall AdS/QCD wave function. In the small-$x$ region, however, these distributions exhibit a divergent behavior as $x\to 0$, which mirrors the behavior of the corresponding gravitational TMDs. The flavor hierarchy remains unchanged in this region as well, with the $u$-quark contribution dominating over the $d$-quark contribution throughout the entire $x$ range. 

The sign change observed in $A_{5}^{(\beta)}(x)$ arises from the interplay of terms in Eq.~\eqref{eq:PDF5}, where the second term becomes dominant in the large-$x$ region and consequently reverses the overall sign of the distribution. At present, unlike  $A_{1}^{(\beta)}(x)$ and $A_{3}^{(\beta)}(x)$, no clear physical interpretation has been established for the gravitational PDF $A_{5}^{(\beta)}(x)$~\cite{Lorce:2023zzg}, although this distribution is closely related to the light-front quark energy density, as it originates from the $\mathcal{T}_q^{+-}$ component of the TMD-EMT, which represents the light-front energy density. In the subsequent section, we demonstrate that the gravitational PDFs $A_{1}^{(\beta)}(x)$ and $A_{3}^{(\beta)}(x)$, obtained from the gravitational TMDs $a_{1}^{(\beta)}(x,k_\perp^{2})$ and $a_{3}^{(\beta)}(x,k_\perp^{2})$ after integrating over the transverse momentum, are directly related to the average longitudinal momentum and the isotropic pressure and shear distribution within the TMD framework
\cite{Lorce:2023zzg}.

\subsection{Mechanical properties of unpolarized proton}

\begin{figure}[htbp]
    \centering
  \includegraphics[width=0.388\linewidth]{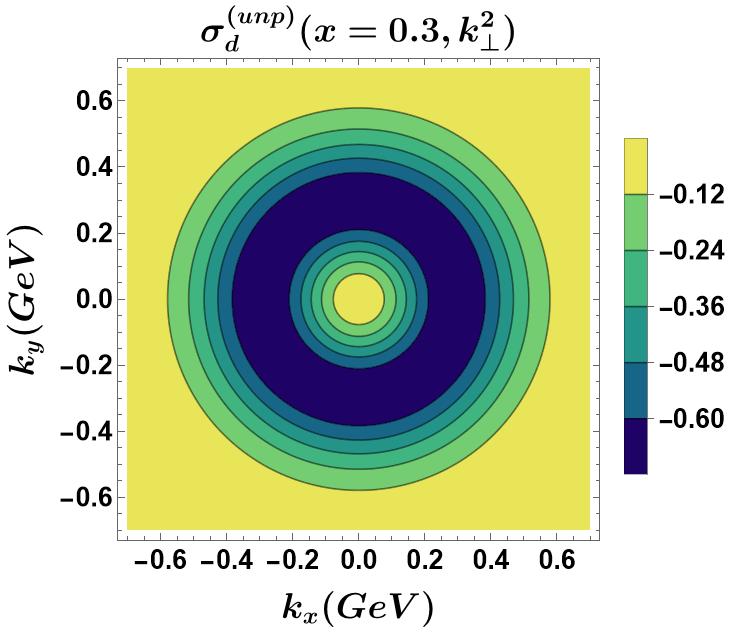}
  \hspace{0.04\linewidth}
    \includegraphics[width=0.388\linewidth]{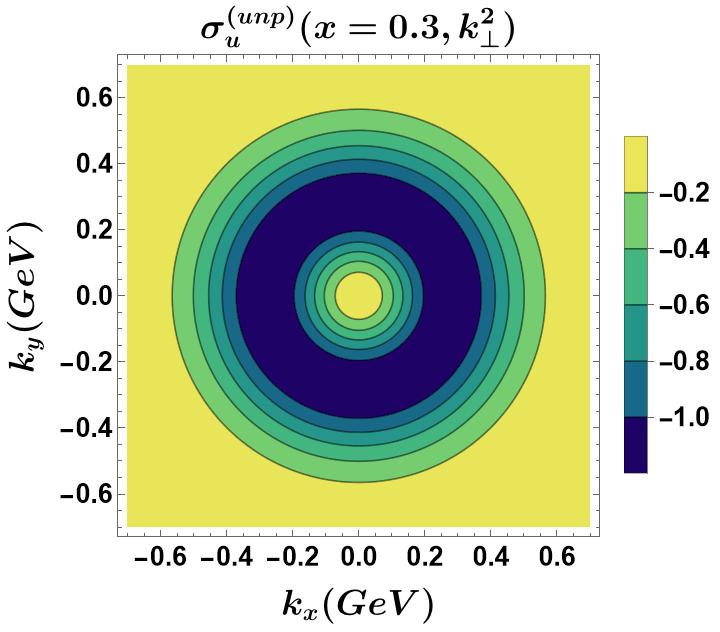}
    \caption{\justifying
   Left panel: Isotropic transverse pressure for the down quark shown as a two-dimensional contour plot in the transverse-momentum ($k_\perp$) plane at fixed momentum fraction $x=0.3$. Right panel: Same as the left panel, but for the up quark. }
    \label{fig:pressurek}
\end{figure}

\begin{figure}[htbp]
    \centering
    \includegraphics[width=0.388\linewidth]{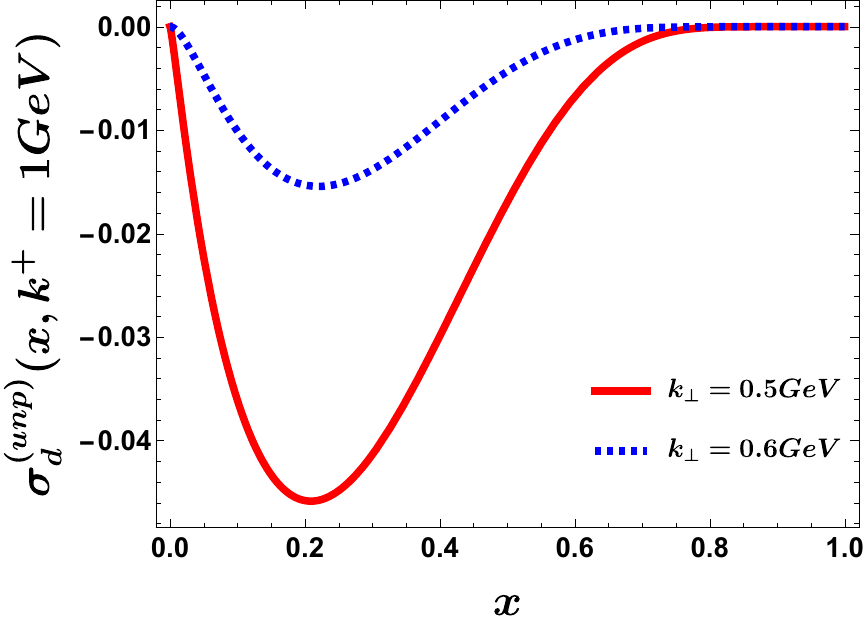}
    \hspace{0.04\linewidth}
    \includegraphics[width=0.388\linewidth]{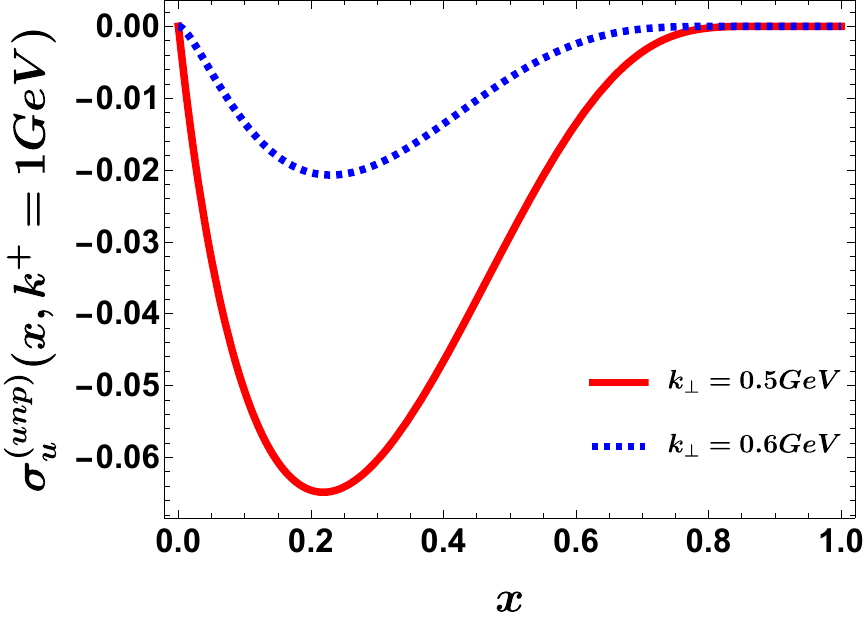}
    \caption{
     Left panel: Isotropic transverse pressure for the down quark shown as a function of the momentum fraction $x$ for different values of the transverse momentum, $k_{\perp}=0.5~\mathrm{GeV}$ and $0.6~\mathrm{GeV}$, at fixed $k^{+}=1~\mathrm{GeV}$. Right panel: Same as the left panel, but for the up quark.
}
    \label{fig:pressurex}
\end{figure}

The transverse components of the energy--momentum tensor, $\mathcal{T}^{ij}$, encode important information about the internal mechanical structure of the system. Following the momentum-space formulation introduced in Ref.~\cite{Lorce:2023zzg}, we decompose the transverse part of the EMT in terms of transverse pressure and shear-force distributions defined directly in transverse-momentum space as
\begin{equation}
\mathcal{T}^{ij}(x,\boldsymbol{k}_\perp)
=
- g_{\perp}^{ij}\, \sigma(x,\boldsymbol{k}_\perp)
+
\left(
\frac{1}{2} g_{\perp}^{ij}
-
\frac{k_{\perp}^i k_{\perp}^j}{\boldsymbol{k}_{\perp}^2}
\right)
\Pi(x,\boldsymbol{k}_\perp)
+
\frac{
k_{\perp}^i \epsilon_{\perp}^{j k_\perp}
+
k_{\perp}^j \epsilon_{\perp}^{i k_\perp}
}{2 \boldsymbol{k}_\perp^2}
\, \Pi_S(x,\boldsymbol{k}_\perp)
+
\epsilon_{\perp}^{ij}\,
\Pi_A(x,\boldsymbol{k}_\perp),
\label{eq:Tij_decomposition}
\end{equation}
where $\sigma(x,\boldsymbol{k}_\perp)$ represents the transverse (isotropic) pressure distribution in momentum space, while $\Pi(x,\boldsymbol{k}_\perp)$ encodes the shear-force (or pressure anisotropy) distribution. The additional structures $\Pi_S(x,\boldsymbol{k}_\perp)$ and $\Pi_A(x,\boldsymbol{k}_\perp)$ correspond to spin-dependent and naive time-reversal-odd contributions that arise specifically in momentum space. The first two terms in Eq.~\eqref{eq:Tij_decomposition} are analogous to the usual decomposition in impact-parameter space obtained by replacing $\boldsymbol{k}_\perp$ with the impact parameter $\boldsymbol{b}_\perp$~\cite{Lorce:2018egm}, whereas the structures proportional to $\Pi_S$ and $\Pi_A$ have no direct counterparts in the impact-parameter-space decomposition.

Following Ref.~\cite{Lorce:2023zzg}, the quark contribution to the 
transverse EMT can be expressed as
\begin{equation}
\mathcal{T}_q^{ij}
=
\frac{1}{P^+}
\left[
\left(
f^{\perp}
-
\frac{M\,\epsilon_{\perp}^{k_{\perp} S_T}}{\boldsymbol{k}_{\perp}^2}\,
f_T^{-}
\right)
k_{\perp}^i k_{\perp}^j
-
\left(
\lambda f_L^{\perp}
+
\frac{M(\boldsymbol{k}_{\perp}\!\cdot\!\boldsymbol{S}_T)}
{\boldsymbol{k}_{\perp}^2}\,
f_T^{+}
\right)
\epsilon_{\perp}^{i k_{\perp}} k_{\perp}^j
\right],
\label{eq:Tij_from_ref}
\end{equation}

By comparing the general tensor decomposition in Eq.~\eqref{eq:Tij_decomposition}
with the explicit quark EMT expression in Eq.~\eqref{eq:Tij_from_ref}, one can
identify the transverse pressure and shear-force distributions. Following the
procedure outlined in Ref.~\cite{Lorce:2023zzg}, we obtain

\begin{equation}
\sigma_q(x,\boldsymbol{k}_\perp)
=
-\frac{1}{2P^+}
\left[
\boldsymbol{k}_{\perp}^2 f^{\perp}
-
M\,\epsilon_{\perp}^{k_{\perp} S_T}\, f_T^{-}
\right],
\qquad
\Pi_q(x,\boldsymbol{k}_\perp)=2\,\sigma_q(x,\boldsymbol{k}_\perp).
\label{eq:Pressure}
\end{equation}

\begin{equation}
\Pi_A^{\,q}
=
\frac{1}{2}\Pi_S^{\,q}
=
-\frac{1}{2P^+}
\left[
\lambda \boldsymbol{k}_{\perp}^2 f_L^{\perp}
+
M(\boldsymbol{k}_{\perp}\!\cdot\!\boldsymbol{S}_T)\,f_T^{+}
\right],
\end{equation}

where $f_{T}^{\pm}=f_{T}\pm \frac{\textbf{k}_{\perp}^2}{2M^2}f_{T}^{\perp}$.

\vspace{0.5em}

In this work, we explicitly determine the isotropic pressure and shear-force distributions for an unpolarized target using the gravitational TMDs obtained in our framework. For an unpolarized proton, Eq.~\eqref{eq:Pressure} can be written as
\begin{equation}
\sigma_{q}^{\text{unp}}
= -\frac{1}{2P^{+}}\,
k_{\perp}^{2}\,
a_{3}^{(\beta)}(x, \textbf{k}_{\perp}^{2}).
\label{eq:unpolarizedpressure}
\end{equation}
Here we have used Eq.~\eqref{eq:model-independent-relation3} to express the pressure in terms of the gravitational TMD $a_{3}^{(\beta)}(x,\boldsymbol{k}_{\perp}^{2})$. 
\vspace{0.3em}

In Figs.~\ref{fig:pressurek}, we present the unpolarized transverse pressure distribution of the proton, defined in Eq.~\eqref{eq:unpolarizedpressure}, as a two-dimensional contour plot in the $k_x$--$k_y$ plane at fixed momentum fraction $x=0.3$ for both up and down quarks in the model. Since the unpolarized isotropic pressure $\sigma_{q}^{\mathrm{unp}}$ depends on the gravitational TMD $a_{3}^{(\beta)}(x,\boldsymbol{k}_\perp^{2})$, which contains a Gaussian factor originating from the soft-wall AdS/QCD wave function, the resulting pressure distribution in the transverse-momentum plane is circularly symmetric for both quark flavors, as expected for an unpolarized proton. The magnitude of the pressure is minimum at the center of the transverse momentum plane; $k_x=k_y=0$. As the transverse momentum increases, the pressure becomes increasingly negative, magnitude being maximum at intermediate $k_\perp$, and gradually approaches zero at large transverse momentum. The negative sign of the pressure indicates a predominantly compressive internal force, consistent with confinement-driven binding dynamics. Quantitatively, the magnitude of the down-quark pressure is smaller than that of the up quark. The up-quark distribution exhibits a steeper maximum in magnitude, signaling a stronger mechanical response in the $u$ sector. This hierarchy reflects the flavor asymmetry in the mechanical structure of the proton within the model~\cite{Maji:2016yqo}. A similar behavior is expected for the isotropic stress distribution defined in Eq.~\eqref{eq:Pressure}, whose magnitude is simply twice that of the corresponding pressure. Therefore, it exhibits the same qualitative features as the plotted pressure distribution, differing only by an overall factor of two.

In Figs.~\ref{fig:pressurex}, we display the same unpolarized transverse pressure distribution as a function of $x$ for different values of the transverse momentum, $k_\perp=0.5~\mathrm{GeV}$(red curves) and $0.6~\mathrm{GeV}$(dashed blue curves), at fixed $k^{+}=1~\mathrm{GeV}$. A notable feature is that the isotropic transverse pressure vanishes at the kinematic endpoints $x\to 0$ and $x\to 1$. In the small-$x$ region, the active quark carries only a negligible fraction of the proton’s longitudinal momentum, and therefore its contribution to the internal mechanical structure becomes insignificant. Conversely, as $x\to 1$, the active quark carries almost the entire longitudinal momentum of the proton, leaving the spectator diquark  with very little momentum; from the plot we see that the pressure in these configurations is strongly suppressed in a bound state.  In the intermediate region $x\sim 0.2$--$0.4$, corresponding to valence-dominated dynamics, the distribution develops a pronounced maximum (negative), indicating the strongest compressive stress in this kinematic domain. The magnitude of the maximum  is larger for larger values of $k_\perp$ as shown in Figs.~\ref{fig:pressurex}, suggesting stronger confining  effects for quarks with large transverse momentum. As $k_\perp$ decreases, the magnitude of the negative pressure decreases, indicating that quarks carrying smaller transverse momentum are less tightly bound. Over the entire $x$ range, the up-quark pressure remains systematically more negative than that of the down quark, further emphasizing the intrinsic flavor asymmetry of the proton’s mechanical structure in this model.

Before concluding this discussion, we briefly comment on the conceptual interpretation of the transverse pressure distribution within the TMD framework. As seen from the plot, the transverse pressure obtained from the gravitational TMD formalism gives the pressure distribution in momentum space, corresponding to the contributions of partons with specified longitudinal momentum fraction $x$ and transverse momentum $\boldsymbol{k}_\perp$, in  contrast to the pressure extracted from gravitational form factors (GFFs) which is in position space\cite{Lorce:2018egm, Polyakov:2018zvc}. In the GFF approach, a Fourier transform with respect to the momentum transfer provides access to spatial distributions of the energy--momentum tensor (EMT), allowing a mechanical interpretation in terms of local pressure and shear forces, as well as the derivation of constraints such as the von~Laue condition, which follows from the local conservation of the EMT. Pressure distributions in two (light-front framework) and three dimensions (Breit frame) have been discussed in the literature.  In contrast, in the case of gravitational TMDs,  there is no access to an average spatial coordinate variable and we get information about the pressure integrated over all space due to partons with specific momentum.

\subsection{Density of longitudinal momentum}

In this subsection, we discuss the physical interpretation of the gravitational TMD $a_{1}^{(\beta)}(x,\mathbf{k}_\perp^{2})$ in terms of the average longitudinal momentum of quarks. The $\mathcal{T}^{++}_q$, component of the TMD energy--momentum tensor  is interpreted as the quark longitudinal momentum density in momentum space~\cite{Lorce:2023zzg}. The average quark longitudinal momentum is therefore obtained by integrating over the quark momentum as
\begin{equation}
\langle k^+ \rangle_q
=
\int dx\, d^2\mathbf{k}_{\perp} \; \mathcal{T}_q^{++}.
\end{equation}
Substituting Eq.~\eqref{eq:A1} into the above expression, we obtain
\begin{equation}
\langle k^+ \rangle_q
=
P^+ \int dx\, d^2\mathbf{k}_{\perp} \; a_{1}^{(\beta)}(x,k_\perp^2).
\label{eq:longitudinal momentum}
\end{equation}

For the numerical evaluation of Eq.~\eqref{eq:longitudinal momentum}, we adopt kinematic conditions relevant to the future Electron–Ion Collider (EIC) by considering a center-of-mass energy of $\sqrt{s_{ep}} = 100~\mathrm{GeV}$. This setup corresponds to an electron beam energy of $10~\mathrm{GeV}$ and a proton beam energy of $250~\mathrm{GeV}$ in the collider frame. In this regime, the proton is ultra-relativistic, and its light-front longitudinal momentum can be approximated as $P^+ \approx 500~\mathrm{GeV}$. 

By numerically integrating Eq.~\eqref{eq:longitudinal momentum} over the longitudinal momentum fraction and transverse momentum of gravitational TMD $a_1^{(\beta)}(x, \textbf{k}_{\perp}^2)$, we obtain the average longitudinal momentum carried by the up quark as
$\langle k^+ \rangle_u = 368.440 \pm 3.404~\mathrm{GeV}$, 
while for the down quark we find 
$\langle k^+ \rangle_d = 169.910^{+0.255}_{-0.244}~\mathrm{GeV}$. 
The larger average longitudinal momentum carried by the up quark compared to the down quark reflects the proton’s valence structure and follows naturally from the flavor decomposition inherent in our model.

\section{conclusion}

In this work, we have taken a first step towards studying the gravitational transverse-momentum-dependent distributions within the light-front quark--diquark model (LFQDM) inspired by the soft-wall AdS/QCD framework, where the proton is described as a bound state of an active quark and a spectator diquark. Within this model, we derived analytical expressions for six unpolarized (T-even) gravitational TMDs and verified that they satisfy the model-independent relations with quark TMDs. By integrating over the transverse momentum, we obtained the corresponding three unpolarized gravitational PDFs. Furthermore, we investigated the mechanical properties of the proton using the gravitational TMDs. In particular, the distribution $a_{3}^{(\beta)}(x,\mathbf{k}_{\perp}^{2})$ is directly connected to the transverse isotropic pressure and shear distributions. The resulting distributions were found to be circularly symmetric in the transverse-momentum plane for both up and down quarks and exhibit negative values, indicating predominantly compressive internal forces consistent with confinement-driven binding dynamics. In addition, we investigated the average longitudinal momentum carried by quarks using the gravitational TMD $a_{1}^{(\beta)}(x,\mathbf{k}_{\perp}^{2})$ and found that the average longitudinal momentum fraction carried by the up quark is larger than that of the down quark, reflecting the flavor asymmetry of the proton within this model.
The gravitational TMDs $a_{1}^{(\beta)}(x,\mathbf{k}_{\perp}^{2})$ and $a_{6}^{(\beta)}(x,\mathbf{k}_{\perp}^{2})$ are related to the twist-2 TMD $f_{1}^{(\beta)}(x,\mathbf{k}_{\perp}^{2})$ and these  can be extracted from experiments. We also presented the pressure distributions in momentum space inside the hadron  obtained from gravitational TMDs in this model. This is conceptually different from the pressure distribution in position space derived using gravitational form factors (GFFs). As a future direction, within the same LFQDM framework we plan to derive analytical expressions for the remaining polarization-dependent gravitational TMDs, including T-odd gravitational TMDs and further investigate the mechanical structure of the proton in momentum space.

\label{sec:conclusion}

\section*{Acknowledgments}
We thank Cédric Lorcé for helpful discussions.

\appendix
\section{Parameterization of TMD-EMT in terms of their component}

In this appendix, we consider specific components of the TMD energy-momentum tensor appearing in Eq.~\eqref{eg: parameterizedTMD}. In particular, by evaluating the components
$\mathcal{T}_q^{++}$, $\mathcal{T}_q^{ij}$, $\mathcal{T}_q^{-+}$, $\mathcal{T}_q^{+i}$, $\mathcal{T}_q^{i+}$, and $\mathcal{T}_q^{-i}$,
we obtain

\begin{equation}
     \mathcal{T}_{q}^{++}(x,\boldsymbol{k_\perp},S)
= P^{+}\left(
a_{1}
- \frac{\epsilon_{\perp}^{\,k_{\perp}S_{T}}}{M}\, a_{1T}^{\perp}
\right),
\label{eq:A1}
 \end{equation}

 \begin{equation}
\begin{aligned}
\mathcal{T}_{q}^{ij}(x,\boldsymbol{k_\perp},S)
=\frac{1}{P^+}\Bigg\{
k_{\perp}^{i} k_{\perp}^{j}\, a_3
-\frac{\epsilon_{\perp}^{k_{\perp} S_T}}{M}
k_{\perp}^{i} k_{\perp}^{j}\, a_{3T}^{\perp}
-\lambda\Big(
k_{\perp}^{i}\,\epsilon_{\perp}^{j k_{\perp}} a_{5L}
+
k_{\perp}^{j}\,\epsilon_T^{i k_{\perp}} a_{6L}
\Big)
- M\Big(
k_{\perp}^{i}\,\epsilon_{\perp}^{j S_T} a_{5T}
+
k_{\perp}^{j}\,\epsilon_{\perp}^{i S_T} a_{6T}
\Big)
\Bigg\},
\end{aligned}
\label{eq:A2}
\end{equation}

\begin{equation}
\mathcal{T}_{q}^{-+}(x,\boldsymbol{k_\perp},S)
=
\frac{M^{2}}{P^{+}}
\left(
a_{1}
+
a_{5}
\right)
-
\frac{M}{P^{+}}\,
\epsilon_{\perp}^{\,k_{\perp} S_{T}}
\left(
 a^{\perp}_{1T}
+
a^{\perp}_{5T}
\right),
\label{eq:A3}
\end{equation}

\begin{equation}
   \mathcal{T}_{q}^{+i}(x,\boldsymbol{k_\perp},S)
=
k_{\perp}^i
\left[
a_6
-
\frac{\epsilon_{\perp}^{k_{\perp} S_{T}}}{M}\,
a^{\perp}_{6T}
\right]-M\epsilon_{\perp}^{iS_{T}}a_{1T}-\lambda \epsilon_{\perp}^{ik_{\perp}}a_{1L},
\label{eq:A4}
\end{equation}

\begin{equation}
\mathcal{T}_{q}^{i+}(x,\boldsymbol{k_\perp},S)
=
k_{\perp}^i\left(
a_7
-
\frac{\epsilon_{\perp}^{k_{\perp} S_T}}{M}\, a^{\perp}_{7T}
\right)
-
M\,\epsilon_{\perp}^{i S_T}\, a_{2T}
-
\lambda\,\epsilon_{\perp}^{i k_{\perp}}\, a_{2L},
\label{eq:A5}
\end{equation}

\begin{equation}
\begin{aligned}
\mathcal{T}_{q}^{-i}(x,\boldsymbol{k_\perp},S)
= \frac{M^2}{(P^+)^2}\Bigg\{
k_{\perp}^i \Bigg[
 a_6 + a_8
-\frac{\epsilon_{\perp}^{k_{\perp}S_T}}{M}
\left( a^{\perp}_{6T} + a^{\perp}_{8T}\right)
\Bigg] - M\,\epsilon_{\perp}^{iS_T}
\left( a_{1T} + a_{3T}\right)- \lambda\,\epsilon_{\perp}^{ik_{\perp}}
\left( a_{1L} + a_{3L}\right)
\Bigg\}.
\end{aligned}
\label{eq:A6}
\end{equation}

Under the flip of the nucleon spin, $S \rightarrow -S$, the unpolarized gravitational TMDs can be extracted using Eqs.~\eqref{eq:A1}--\eqref{eq:A6} as

\begin{equation}
    \mathcal{T}_{q}^{++}(x,\boldsymbol{k_\perp},+) + \mathcal{T}_{q}^{++}(x,\boldsymbol{k_\perp},-)
= 2 P^{+} \, a_{1}(x,\boldsymbol{k_\perp^2})
\label{eg:a1}
\end{equation}

\begin{equation}
\mathcal{T}_{q}^{ij}(x,\boldsymbol{k_\perp},+)
+ \mathcal{T}_{q}^{ij}(x,\boldsymbol{k_\perp},-)
= 2\,\frac{k_{\perp}^{i} k_{\perp}^{j}}{P^{+}}a_{3}(x,\boldsymbol{k_\perp^2})
\label{eg:a3}
\end{equation}

\begin{equation}
     \mathcal{T}_{q}^{+-}(x,\boldsymbol{k_\perp},+) + \mathcal{T}_{q}^{+-}(x,\boldsymbol{k_\perp},-)
= \frac{2M^2}{P^+} \bigg(a_{1}(x,\boldsymbol{k_\perp^2})+a_{5}(x,\boldsymbol{k_\perp^2})\bigg)
\label{eg:a5}
\end{equation}

\begin{equation}
     \mathcal{T}_{q}^{+i}(x,\boldsymbol{k_\perp},+) + \mathcal{T}_{q}^{+i}(x,\boldsymbol{k_\perp},-)
= 2 \textbf{k}_{\perp}^i \, a_{6}(x,\boldsymbol{k_\perp^2})
\label{eg: a6}
\end{equation}

\begin{equation}
     \mathcal{T}_{q}^{i+}(x,\boldsymbol{k_\perp},+) + \mathcal{T}_{q}^{i+}(x,\boldsymbol{k_\perp},-)
= 2 \textbf{k}_{\perp}^i \, a_{7}(x,\boldsymbol{k_\perp^2})
\label{eg:a7}
\end{equation}

\begin{equation}
     \mathcal{T}_{q}^{-i}(x,\boldsymbol{k_\perp},+) + \mathcal{T}_{q}^{-i}(x,\boldsymbol{k_\perp},-)
= 2\bigg(\frac{M}{P^+}\bigg)^2 \bigg(a_{6}(x,\boldsymbol{k_\perp^2})+a_{8}(x,\boldsymbol{k_\perp^2})\bigg)
\label{eg:a8}
\end{equation}

\bibliography{ref.bib}

\end{document}